\documentclass{aa}
\usepackage{graphicx}
\usepackage{txfonts}
\usepackage{natbib}
\usepackage{longtable}
\usepackage{scrextend}
\usepackage[shortlabels]{enumitem}
\usepackage{siunitx}

\newcommand{\ageGyr}{\Bigl (\frac{\tau}{1{\rm Gyr}}\Bigr )}

\usepackage{hyperref}	
\usepackage{amsmath}    

\usepackage{orcidlink}
\usepackage{xspace}
\newcommand{\zmax}{\ensuremath{|z_{\textup{max}}|}\xspace}

\hypersetup{colorlinks=true,linkcolor=blue,citecolor=blue,filecolor=blue,urlcolor=blue}
\begin{document}





\title{Disc dichotomy signature in the vertical distribution of [Mg/Fe]  \\
and the delayed gas infall scenario}
\author { E. Spitoni \,\orcidlink{0000-0001-9715-5727} \inst{1,2,3}  \thanks {Email to: emanuele.spitoni@oca.eu},
  V. Aguirre B\o rsen-Koch\,\orcidlink{0000-0002-6137-903X}   \inst{3}, K. Verma \,\orcidlink{0000-0003-0970-6440} \inst{4,3} \and A. Stokholm  \,\orcidlink{0000-0002-5496-365X}\inst{5,6,3} }
  \institute{Universit\'e C\^ote d'Azur, Observatoire de la C\^ote d'Azur, CNRS, Laboratoire Lagrange, Bd de l'Observatoire,  CS 34229, 06304 Nice cedex 4, France \and Konkoly Observatory, Research Centre for Astronomy and Earth Sciences, Konkoly Thege Mikl\'{o}s \'{u}t 15-17, H-1121 Budapest, Hungary \and Stellar Astrophysics Centre, Department of Physics and
  Astronomy, Aarhus University, Ny Munkegade 120, DK-8000 Aarhus C,
  Denmark  
  \and Instituto de Ciencias del Espacio (ICE, CSIC), Campus UAB, Carrer de Can Magrans, s/n, 08193 Cerdanyola del Valles, Spain
  \and Dipartimento di Fisica e Astronomia, Universit\`a degli Studi di Bologna, Via Gobetti 93/2, I-40129 Bologna, Italy \and INAF -- Osservatorio di Astrofisica e Scienza dello Spazio di Bologna, Via Gobetti 93/3, I-40129 Bologna, Italy}

\date{Received xxxx / Accepted xxxx}
\abstract
{The analysis of  the  Apache Point Observatory Galactic Evolution
Experiment project (APOGEE)  data suggests the existence of a clear distinction between two sequences of disc stars  in the [$\alpha$/Fe] versus [Fe/H] abundance ratio space: known as the high- and low-$\alpha$ sequence, respectively. This dichotomy  also emerges  from the analysis  of the  vertical distribution of the [$\alpha$/Fe] abundance ratio.
}
{We aim to test whether  the revised  two-infall chemical evolution models designed to reproduce the low- and high-$\alpha$ sequences in the  [$\alpha$/Fe] versus [Fe/H]  ratios  in the solar neighbourhood are also capable to  predict the  disc bimodality  observed in the  vertical distribution of  [Mg/Fe]  in APOGEE DR16 data.}
{Along with  the chemical composition of  simple stellar populations born at different Galactic times   predicted by our reference chemical evolution models in the solar vicinity, we provide their maximum vertical height above the Galactic plane \zmax computed  assuming the  relation between the vertical action and stellar age in APOGEE thin disc stars. }
{The vertical distribution of the [Mg/Fe] abundance ratio predicted by the reference chemical evolution models  is in agreement with the one observed   combining the APOGEE  DR16 data (chemical abundances)   and the astroNN catalogue (stellar ages, orbital parameters) for stars younger than 8 Gyr (only low-$\alpha$ sequence stars). Including the high-$\alpha$ disc component, the dichotomy   in the vertical [Mg/Fe] abundance distribution is  reproduced considering the  observational cut in the Galactic height of $|z|<2$ kpc. However, our model predicts a too  flat  (almost constant)  growth  of the  maximum vertical height \zmax quantity  as a function of [Mg/Fe]  for high-$\alpha$ objects in contrast with the median values from APOGEE data. Possible explanations for such a tension  are: i) the APOGEE sample with $|z|<2$ kpc is more likely  contaminated by  halo stars, causing the median values to be kinematically hotter, ii)  external perturbations such as minor mergers that the Milky Way experienced in the past could have heated up the disc, and  the heating of the orbits cannot  be modeled  by only scattering processes.  Assuming  for APOGEE-DR16  stars ($|z|<2$~kpc) a  disc dissection  based on chemistry,  the  observed  \zmax distributions for  high-$\alpha$ and low-$\alpha$ sequences are in good agreement with our model predictions if we consider in the calculation the  errors in the vertical action estimates.
Moreover, a better agreement between   predicted  and observed stellar distributions at different Galactic vertical heights is achieved if asteroseismic ages are included as a constraint in the best-fit model calculations.
}
 { The signature of a delayed  gas infall episode which gives  rise to  a hiatus in the star formation  history of the Galaxy are imprinted both in   the  [Mg/Fe] versus [Fe/H] relation  and in vertical  distribution of [Mg/Fe] abundances in the solar vicinity.}


\titlerunning{Vertical  distribution of [Mg/Fe]}

\authorrunning{Spitoni et al.}

\maketitle

\section{Introduction}

The analysis of the Apache Point Observatory Galactic Evolution Experiment project (APOGEE) data \citep{Nidever:2014fj, hayden2015,Ahumada2019,queiroz2020,vincenzo2021} highlighted the presence of  two distinct sequences  in the [$\alpha$/Fe] versus [Fe/H] abundance ratio space for disc stars: 
the so-called high-$\alpha$ sequence, classically associated with an old population of stars in the thick disc, and the low-$\alpha$ sequence, which mostly comprises of relatively young stars in the thin disc.
 This dissection has been also revealed by other observational campaigns:  the Gaia-ESO  survey \citep[e.g.,][]{RecioBlanco:2014dd,RojasArriagada:2016eq,RojasArriagada:2017ka} and the Arch\'eologie avec Matisse Bas\'ee sur les aRchives de l’ESO  project (AMBRE;  \citealt{Mikolaitis:2017gd,santosperal2021}), the Galactic Archaeology with HERMES survey (GALAH; \citealt{buder2019, buder2021}) and  the Large sky Area Multi Object fiber Spectroscopic Telescope (LAMOST; \citealt{zheng2021}).

  Cosmological hydrodynamic simulations of Milky Way-like galaxies, \citep{kobayashi2011, snaith2016} predict such a bimodality in the  distribution of chemical elements.  While in  \citet{vincenzo2020} this dichotomy  is mainly attributed to the interplay between  infall and ouflow events,
in the dynamical model presented by \citet{clarke2019} it arises from the fragmentation of  the early gas-rich disc.
 On the other hand, in several theoretical models of Galactic disc evolution, it has been proposed that this bimodality is strictly connected to a delayed gas-accretion episode of primordial composition
 \citep[i.e.][]{noguchi2018, buck2020,lian2020,Khoperskov2020, agertz2021}.
Moreover, the AURIGA simulations presented by \citet{grand2018}  clearly point out that a bimodal distribution in the [Fe/H]-[$\alpha$/Fe] plane is a consequence of a significantly lowered gas accretion rate at ages between 6 and 9 Gyr.  \citet{verma2021}  compared AURIGA simulations
  with 7000 stars with asteroseismic, spectroscopic, and astrometric data available, and concluded that
    the emerged abundance dichotomies in the  [$\alpha$/Fe] versus [Fe/H]   plane   look qualitatively similar to observations.

\citet{spitoni2019}  and \citet[][herafter ES20]{spitoni2020} revised the classical two-infall  chemical evolution model \citep{chiappini1997}  in order to  reproduce  
 APOKASC (APOGEE+ Kepler Asteroseismology Science Consortium, \citealt{victor2018}) sample. They  invoked the presence of a delayed  gas infall episode ($\sim$ 4 Gyr) in order to reproduce the high- and low-$\alpha$ sequence stars including  also precise asteroseismic ages as constraint.
This delayed infall of gas gives rise to the low-$\alpha$ sequence by bringing pristine metal-poor gas into the system which dilutes the metallicity of interstellar medium while keeping [$\alpha$/Fe] abundance almost unchanged. 
Similarly, also in  \citet[][herafter ES21]{spitoni2021} a significant delay between the two gas infall episodes is  fundamental  in order  to  reproduce APOGEE DR16 data in the solar vicinity, whereas in  the innermost regions a chemical enriched gas infall is required in order to reproduce observed [Mg/Fe] versus [Fe/H] ratios   as also suggested by \citet{palla2020}.

 An inside-out formation of the thin disc of the Galaxy naturally emerges from the  multi-zone chemical evolution model of ES21, i.e. the  Milky Way formed on much shorter timescales in the inner than the outer regions. Such a mechanism has been found also in complex cosmological simulations of galaxy formation \citep{brook2012,bird2013,kobayashi2011,vincenzo2020}.
As underlined by \citet{bird2013},  the growth of the  simulated galaxy also  follows  an “upside-down” evolution in the vertical direction, namely old stars form in a relatively thick component and are kinematically heated very quickly after their birth. Later on,  low-$\alpha$  stellar populations form in successively thinner disc.
In principle, this is in agreement with the assumptions of the two-infall model, shorter time-scales of gas accretion characterise the formation of the thick disc. 
However, the above-mentioned  ES20 and ES21 models  could make predictions just on  projected quantities on the Galactic plane. 
Our principal aim with this work is  to study  the vertical distribution of chemical elements assuming  simplified dynamical prescriptions using  these chemical evolution models.

 In fact, in order to better understand the processes that dominated the formation and evolution of  the Galactic disc, it is crucial to compare  model predictions with  also the observed  vertical   [$\alpha$/Fe] distribution of stars at different heights above the Galactic  plane. 
In the past,  the vertical abundance gradients  have been  subject of several investigations. For instance, 
\citet{Schlesinger2014} analysed G dwarfs from the Sloan Extension for Galactic Understanding and Exploration (SEGUE) survey,  and  showed the presence of negligible   vertical metallicity gradient in the Milky Way’s disc
for different [$\alpha$/Fe] subsamples.
It suggests that stars formed in different epochs shared  similar star formation processes and evolution.
\citet{mikolaitis2014} used the spectra of around 2000 FGK dwarfs and giants from the Gaia-ESO survey  iDR1, and
found that thick disc stars show a shallower vertical metallicity gradient than the thin disc, an [$\alpha$/Fe] ratio gradient in the opposite sense than that of the thin disc, and positive vertical individual [$\alpha$/M] and [Al/M] gradients  (where M is the metallicity).
\citet{doung2018},
using data from the GALAH  survey, determined the vertical properties of the Galactic thin and
thick discs near the solar neighbourhood. 
The median [$\alpha$/M] increases as a function of height, as noted previously
by \citet{Schlesinger2014} and \citet{mikolaitis2014}. However, unlike the metallicity, they find that the $\alpha$-abundance profile
does not vary smoothly with $|z|$.

  Beside the chemical signatures, the orbital properties of stars and in particular the change of dynamical actions over time could provide important constraints on the main evolutionary processes that have determined stellar redistribution. 
In \citet{beane2018} and \citet{ness2019}  the connections between dynamical actions, ages, and chemical abundances in disc stars were discussed. 
In particular, \citet{gandhi2019} found  that at all ages, the high- and low-$\alpha$ sequences are dynamically distinct and that selections in the actions space can provide  an efficient method to separate distinct dynamical populations.

More recently, \citet{vincenzo2021}  highlighted   the presence of   bimodality in the  vertical [$\alpha$/Fe] distribution of  APOGEE DR16 that can be well modelled adopting  a double Gaussian stellar distribution:
one component describing the low-$\alpha$ population with scale height $z_1=0.45$ kpc and one describing the high-$\alpha$ population with scale-height $z_2=0.95$ kpc.

In  this article, we compare the vertical structure of the  [Mg/Fe] distribution as emerging from APOGEE DR16 data in the solar neighborhood  with   the results of  revised  two-infall chemical evolution models presented by ES20 and ES21. To do so, besides   the predicted chemical composition of  simple stellar populations (SSPs,  defined as an assembly of coeval and chemically homogeneous stars) formed at different Galactic times,  we also provide  the respective orbital parameters (i.e. maximum vertical height above the Galactic plane \zmax) computed  using the  recent relation between the vertical action $J_z$  and stellar age found for APOGEE thin disc stars \citep{ting}. The main  scope of  this work is to confirm that the disc bimodality found in APOGEE DR16 stars in the [$\alpha$/Fe] versus Galactic vertical height \citep{vincenzo2021} can also be interpreted as the signature of a delayed accretion of gas  which  happened  about $\sim 4.3$ Gyr  after the Galactic formation.

The paper is organised as follows. 
In Section \ref{data:s}, the observational data  will be presented.  In Section \ref{s:chemmod}, we draw the main characteristics of the reference chemical evolution models for the high- and low-$\alpha$ adopted in  this study.
In Section \ref{s:jz}, 
we describe the adopted methodology to compute \zmax for stars born at different Galactic times and the vertical [Mg/Fe] gradient. In Section \ref{s:results}, we present our results and finally, in Section \ref{s:conclusions}, we draw our conclusions.

\section{APOGEE DR16 data and astroNN catalogue}
\label{data:s}
In this study, we consider  Mg and Fe abundances  provided by APOGEE DR16 \citep{Ahumada2019}  for investigating the region with  Galactocentric distances between  6 and 10 kpc as computed  by \citet{leung2019} and reported  in the value-added catalogue  astroNN\footnote{\href{https://data.sdss.org/sas/dr16/apogee/vac/apogee-astronn}{https://data.sdss.org/sas/dr16/apogee/vac/apogee-astronn}} 
catalogue.
astroNN is a deep-learning software, which was applied to APOGEE DR16 spectra in order to determine stellar parameters,  distances  \citep{leung2019}, and ages \citep{mackereth2019}. In addition, the above-mentioned catalogue includes some of the most important orbital properties for stars (i.e. eccentricities, peri/apocenter radii, maximal disc height \zmax, orbital actions, frequencies and angles) computed by \citet{mackereth2018} assuming the  {\tt MWPotential2014}  gravitational potential from \citet{bovy2015}.

Following ES21, we choose only stars that are part of the Galactic disc with the same quality cuts suggested in \citet{weinberg2019} assuming signal-to-noise ratio $(S/N) >80$,  logarithm of surface gravity  between $1.0< \log g<2.0$.

\begin{figure}
\begin{centering}
\includegraphics[scale=0.7]{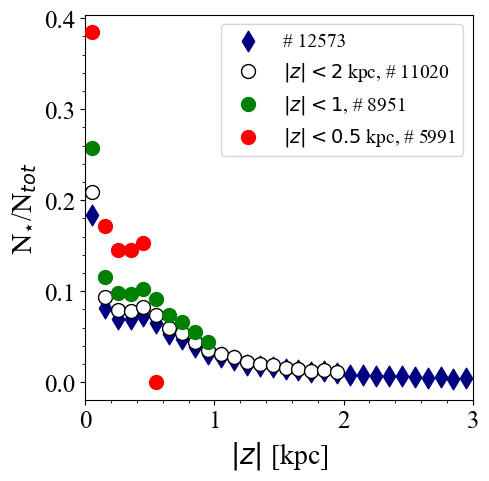}
\caption{  The normalised distributions  of   the vertical height $|z|$  above the Galactic plane  for APOGEE-DR16 adopting the selection cuts presented in Section \ref{data:s}. White, green and red circles stand for distributions assuming   $|z|<2$ kpc and $|z|<1$ kpc, $|z|<0.5$ kpc, respectively.  The case without any cut in the vertical height is labeled with the blue diamonds. The respective total number of stars are also indicated. 
}
\label{dist_absz}
\end{centering}
\end{figure}

In this study, we analyse  different regions above and below the Galactic planes, considering stars with the following observed vertical heights: $|z|<0.5$ kpc, $|z|< 1$ kpc, and $|z|< 2$ kpc in order to  better analyse the vertical structure. 
 In Fig. \ref{dist_absz}, we note that the vertical distribution with $|z|< 2$ kpc  resembles well the behaviour of the whole APOGEE-DR16 stellar sample in the region with Galactocentric distances enclosed between 6 and 10 kpc.

 As underlined by \citet{weinberg2019}, the    adopted selection in  $\log g$ leaves only stars on the upper red giant branch (RGB), i.e. the most luminous ones. It ensures that the stars in our sample can be observed by APOGEE over most of the distance range considered in this study, minimizing distance-dependent changes in the population being analysed.

\section{Chemical evolution models for solar vicinity}\label{s:chemmod}
In this Section, we introduce the main  assumptions  of the two-infall chemical evolution model  proposed by ES20 and ES21, providing  a few details on  the parametrization of the  infall and star formation. 
In both models, the Milky Way disc is assumed to be formed by two distinct 
episodes of gas accretion onto the Galactic plane (i.e. with null vertical height, $z=0$ kpc). The gas infall rate is expressed  as  follows:
\begin{equation}
\mathcal{I}_i(t,z=0) \equiv (\mathcal{X}_i)_{inf} \left[ \mathcal{A}_1 \, e^{-t/ t_{1}}+ \theta(t-t_{{\rm max}}) \mathcal{A}_2 \, e^{-(t-t_{{\rm max}})/ t_{2}} \right],
\label{a}
\end{equation}
where $t_{1}$ and $t_{2}$ are the infall time-scales for the thick and thin disc components, respectively. The coefficient $t_{\rm max}$ indicates the delay of the beginning of the second infall, hence
the time for the maximum accretion rate on the second infall episode.
Finally, the quantities  $\mathcal{A}_1$ and $\mathcal{A}_2$ are obtained imposing a fit to the observed current total surface mass density in the solar neighbourhood.
As for the total surface density in the solar neighborhood    we adopted the value of   47.1 $\pm$ 3.4 M$_{\odot} \mbox{ pc}^{-2}$ suggested by 
 \citet{mckee2015}.
  We stress that   these chemical evolution models do not make any assumption about the vertical growth of the gaseous and stellar discs.

  In both ES20 and ES21 models the  numerical treatment is based on Matteucci´s chemical evolution code  (all the details could be retrieved in  the review  \citealt{matteucci2021} and in the book \citealt{matteucci2012}).  The star formation rate (SFR) is expressed as the \citet{kenni1998} law,
$\psi\propto \nu \sigma_{g}^{k}$,
where $\sigma_g$ is the gas surface
 density, and $k = 1.5$ is the exponent. 
The quantity $\nu$ is the star formation efficiency (SFE) and it is fixed to the values of $\nu_1=2$ Gyr$^{-1}$ and $\nu_2=1$ Gyr$^{-1}$  for the high-$\alpha$ and
low-$\alpha$ sequences, respectively (see Table \ref{tab_mcmc}).
The type Ia SN rate has been computed following \citet{greggio1983}  and \citet{matteucci1986} prescriptions (see \citealt{spitoni2009} for the rate expression).

 Although  the disc bimodality in the  [$\alpha$/Fe] versus [Fe/H] space in a simulated Milky Way-like galaxy in the cosmological framework of \citet{vincenzo2020}  is attributed to the interplay between  infall and ouflow events, in ES20 and ES21 no Galactic winds have been considered.  This choice has been motivated by \citet{melioli2008, melioli2009} and   \citet{spitoni2008,spitoni2009} studies  on the Galactic fountains (i.e., processes originated by the explosions of Type II SNe in OB associations). They found that the ejected metals fall back close to the same Galactocentric region where they are delivered and thus do not modify significantly the chemical evolution of the disc as a whole.

The adopted nucleosynthesis prescriptions
are the ones suggested by \citet{francois2004}. 
We briefly recall here the main assumptions for  the  chemical elements analysed in our study (i.e., Mg and Fe).
The authors modified the Mg yields from massive stars from \citet{WW1995} to reproduce the solar  abundance value.
 Mg yields from stars in the range 11-20 M$_ {\odot}$ have been increased by a factor of  7  whereas those from stars in the mass range 20 M$_ {\odot}<$M $<$ 100 M$_ {\odot}$ are lower than predicted by \citet{WW1995} by a factor of 2 on average.   No modifications are needed for the yields of  Fe, as computed for solar chemical composition.  
 The complete grid of the modified  yields  can be retrieved from Table 1 of \citet{francois2004}.

Concerning Type Ia SNe, 
the theoretical Mg yields by \citet{iwamoto1999} have been increased  by a factor of 5, in order 
 to  preserve the observed pattern of [Mg/Fe] versus [Fe/H].
  The assumed IMF contains much less massive stars than other works in the community, and this affects the Mg enrichment, requiring more Mg from Type Ia SNe. 
The prescription
for single low- and intermediate- mass stars is from \citet{van1997}, for the case of the mass loss parameter 
which varies with metallicity \citep[see][model5]{chiappini2003}.

  The choice of such  ad-hoc  nucleosynthesis  prescriptions is supported by the fact that   stellar yields are  still a relatively uncertain component of chemical evolution models \citep{romano2010}.  
This set of yields  has been   widely used in the past by Matteucci’s group in Trieste (see \citealt{matteucci2021})
  and turned out to be able to reproduce the main features of the solar neighbourhood \citep[e.g.,][]{cescutti2007,spitoni2011, Mott2013, spitoni2014,spitoni2017,
 spitoni2015, spitoni2D2018, spitoni2019,vincenzo2019}.
   Updated  nucleosynthesis prescriptions for massive stars in  the galactic chemical evolution of Mg still present problems in reproducing the stellar data.  For instance,   \citet{prantzos2018}  presented chemical evolution  with metallicity-dependent weighted rotational velocities  by \citet{chieffi2013} but, as adopting the yields of \citet{WW1995},  the evolution of Mg is not well reproduced. 
 In \citet{kobayashi2020}, using the coalescence's for massive stars from \citet{kobayashi2011} and in presence of failed SNe, they were able to reproduce the [Mg/Fe] versus [Fe/H] in the solar neighbourhood.  \citet{cote2017} studying the chemical evolution of Sculptor  using the NuGrid stellar yields \citep{ritter2018}, highlighted that the model results underestimated the observed  [Mg/Fe] versus [Fe/H] relation. 
 
  In order to be consistent with \citet{francois2004} prescriptions, the initial stellar mass function (IMF) formalised by \citet{scalo1986}   is adopted and assumed constant in time and space. 
 More recent IMF formulations has been assumed in the chemical  evolution model of \citet{kobayashi2020}  (\citealt{kroupa2008} IMF),   \citet{prantzos2018} (\citealt{kroupa2002} IMF) but still  high-mass end is highly uncertain and affected by large systematic uncertainty (also due to binary fraction).

A Bayesian framework based on MCMC methods\footnote{The affine invariant MCMC ensemble sampler code, "\textit{emcee}: the mcmc hammer", developed by  \citet{goodman2010,foreman} was used to  sample the posterior probability distribution. }
was used in  ES21 to fit the APOGEE  DR16 chemical abundance ratios at different Galactocentric distances, whereas models in ES20  were constrained by both chemical abundances and stellar ages (APOKASC sample). In both cases, the free parameters of the model were: the infall time-scales $t_1$ and $t_2$, present-day total surface mass density ratio  $\sigma_2/\sigma_1$ between the  low- and high-$\alpha$ sequences, and the delay $t_{\rm max}$. More details on these parameters  are provided  in the next Sections, \ref{S20} and \ref{S21}.

\subsection{The reference model from ES20}
\label{S20}
In  ES20,
 the presented chemical evolution models have been  designed to fit the observed chemical abundance ratios and asteroseismic ages of the APOKASC stars \citep{victor2018}. This sample had about 1200 red giants from an annular region with a width of 2~kpc in the solar vicinity.  The stellar properties for this sample were determined by fitting the photometric, spectroscopic, and asteroseismic observables using the BAyesian STellar Algorithm code \citep[{\tt BASTA};][]{silvaaguirre2015,silvaaguirre2017,aguirre2021}.  
 In Table \ref{tab_mcmc}, we report the best-fit model parameters as predicted by the MCMC calculation performed  with $\nu_1=2$ Gyr$^{-1}$ and $\nu_2=1$ Gyr$^{-1}$ (case \emph{M2} in their paper).
In agreement with  the classical two-infall model by \citet{chiappini1997}, the first gas infall is characterised
by a short timescale of accretion ($t_1<<t_2$).
Furthermore, we notice the presence of a significant delay $t_{\rm max}$ between the two infall accretion episodes  ($\sim$ 4.6 Gyr), as originally found by \citet{spitoni2019} but without performing any quantitative analysis for parameter estimation. 
  An additional  observational evidence supporting this scenario has been presented by \citet{nissen2020}  analysing the High Accuracy Radial velocity Planet Searcher (HARPS) spectra of local solar twin stars. They found that the age-metallicity distribution shows the presence of  two diverse populations characterised  by a clear age separation. The authors suggested that these two sequences may be interpreted as evidence of two episodes of accretion of gas onto the Galactic disc with quenching of star formation in between them.

\subsection{The reference model from ES21}
\label{S21}
In ES21, they presented  a multi-zone two-infall chemical evolution model with  quantitatively inferred  free parameters by fitting the APOGEE DR16 \citep{Ahumada2019} abundance ratios at different Galactocentric distances.
In particular, the model computed at 8 kpc has been constrained by the  [Mg/Fe] and [Fe/H]  ratios of about 9200 stars  located in the  annular region enclosed between 6 and 10 kpc and  vertical height $|z|$ < 1 kpc.
A significant difference between this model and that of ES20  emerges from Table \ref{tab_mcmc}; the shorter  timescales $t_1$ and $t_2$  are predicted by ES21 model compared to ES20 model. This is  due to the fact that  stellar ages from asteroseismology  were not available for constraining the chemical evolution models in the case of ES21.
The predicted   present-day total surface mass density ratio   between the  low- and high-$\alpha$ sequences of $\sigma_2/\sigma_1=5.635^{+0.214}_{-0.162}$, is in very good agreement  with the ratio derived by \citet{fuhr2017} for  local mass density ratio  of $5.26$.  The presence of an important delay $t_{\rm max}$ is confirmed also in this case as shown in Table \ref{tab_mcmc}.  In addition, the model reproduces important observational constraints of the whole Galactic disc, such as the present-day [Mg/H] abundance gradient, the  profiles of the SFR, the stellar and gas surface densities radial distributions (see Section 4.3 of \citealt{spitoni2021} for a discussion on the global properties of the Galactic disc reproduced by ES21 model).

\begin{table}
\begin{center}
\tiny
\caption{Summary of the main parameters of ES21 (computed at 8 kpc) and ES20 (case \emph{M2})  models. We show the 
 best-fit accretion timescales ($t_1$ and  $t_2$), 
the present-day total surface mass density ratio ($\sigma_{2}$/ $\sigma_{1}$) and delay $t_{{\rm max}}$ predicted by the MCMC calculations.  
}
\label{tab_mcmc}
\begin{tabular}{c|cc}
\hline
  \hline
 &  \multicolumn{2}{c}{\it Models}  \\

  & ES20& ES21 \\

  \hline
 & &\\
 $t_{1}$ [Gyr]&1.264$^{+0.119}_{-0.090}$&0.103$^{+0.007}_{-0.006}$ \\
   & &\\

$t_{2}$ [Gyr]&11.282$^{+0.954}_{-0.943}$&4.110$^{+0.145}_{-0.127}$\\
 & &\\

 $\sigma_{2}$/ $\sigma_{1}$&4.176$^{+0.167}_{-0.178}$&5.635$^{+0.214}_{-0.162}$  \\
 & &\\
 $t_{\rm max}$ [Gyr]&4.624$^{+0.135}_{-0.099}$&4.085$^{+0.021}_{-0.032}$ \\
 & &\\
 \hline
\end{tabular}
\end{center}
\end{table}

\section{Stellar Orbital properties: the vertical action, $J_z$, and the maximum vertical height, \zmax}
\label{s:jz}
First in  Section \ref{tings}, we briefly recall the relation between the average vertical action  and the stellar ages proposed by \citet{ting}. In Section  \ref{cons}, we discuss the conservation of the vertical action in steady-state potentials and in  ``real''  late-type galaxies. Finally, in Section \ref{zmax_c} we compute the  maximum vertical excursion \zmax from the Galactic midplane \zmax as a function of the stellar ages using both the \citet{ting} relation and the conservation of the vertical action.

\subsection{\citet{ting} relation for solar vicinity}
\label{tings}
The action-angle variables are useful quantities to describe the evolution of stellar orbits in steady-state potentials. 
In particular, here we are interested in  the vertical action  $J_z$, defined as:
\begin{equation}\label{eq:action}
  J_z \equiv \frac{1}{2\pi}\int {\rm d}z\; {\rm d}v_z = \frac{1}{2\pi}\oint {\rm d}z\; v_z,
\end{equation}
where $z$ and $v_z$ indicate the vertical height and velocity  along the orbit, respectively.
The vertical action thus quantifies the movement along the $z$ direction of the orbit of a star. These actions are integrals of motions and thus in a static, axis-symmetric potential, they are conserved quantities (see the discussion in next Section).
 \citet{ting} presented a parametric model for the variation of vertical action $J_z$ distribution (i.e. global vertical temperature) analysing  the subsample of APOGEE  red clump stars in the Galactic disc  \citep{ting2018}  with proper motions from the Gaia DR2  \citep{gaia2_2018}.
In \citet{ting}, ages have been computed establishing  an empirical, high-dimensional mapping from
the APOGEE normalised spectra to stellar ages via a fully
connected neural network.

Because red clump stars   are strongly biased against stars older than 8 Gyr, 
\citet{ting} considered only thin disc stars as a function of the age $\tau$, stars younger than 8 Gyr in their model. They claimed the existence of a neat relation between the vertical action and the stellar age, i.e. $J_z$ versus $\tau$.
 \citet{ting} interpreted the distribution of the vertical actions as  combination of the vertical action at birth (i.e. vertical birth temperature) which characterised stars when formed  out of the Galactic disc, plus the subsequent heating  (defined as an increase in $J_z$) that occurred  subsequently).
This relation computed  at the solar radius can be written as: 

\begin{equation}
\label{eq_ting}
\widehat{J_z}(R_{\odot}, \tau) = 0.91+1.81 \cdot \ageGyr^{1.09}\, \mbox{    [kpc km s}^{-1}],
\end{equation}
\noindent
where on the right side the factor $0.91$ is associated with vertical birth temperature,
and the second term indicates the age dependence of the vertical action (after birth) which follows a power law of exponent $\sim 1$.
In Eq. (\ref{eq_ting}), we label the vertical action with $\widehat{J_z}$ instead of $J_z$ to stress that the proposed model is designed to fit the global observed relation for stellar ages  and it should not be interpreted as the temporal evolution of the vertical action $J_z$ of a single stellar population. 

This relation suggests that the vertical heating of all stellar populations is dominated by orbit scattering  but the  scattering amplitude varies with the Galactic epoch. \citet{ting} showed that a model  with an exponential decrease in the SFR and an inside-out growth of the stellar disc can reproduce the range of power-law indices.

\begin{figure}
\begin{centering}
\includegraphics[scale=0.48]{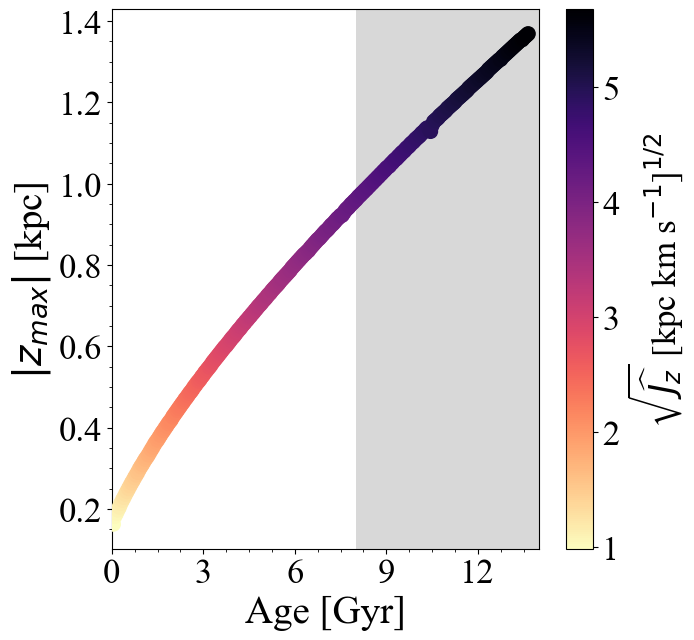}
\caption{Maximum vertical excursion \zmax   predicted for sequential SSPs  born at 8 kpc at different Galactic evolutionary times  as a function of their ages adopting  Eq. (\ref{eq_ting}) and imposing the conservation of $J_z$ in the  orbital integration (see Section  \ref{zmax_c}). The colour coding indicates the values of the $\sqrt{J_z}$ quantity.
We recall that Eq. (\ref{eq_ting})  has been retrieved for thin disc stars in the solar vicinity  for  APOGEE stars younger than 8 Gyr.
The shaded region shows SSPs with ages larger than 8 Gyr, for which the use of Eq. (\ref{eq_ting}) may not be justified. 
}
\label{jz_zmax}
\end{centering}
\end{figure}

\begin{figure*}
\begin{centering}
\includegraphics[scale=0.42]{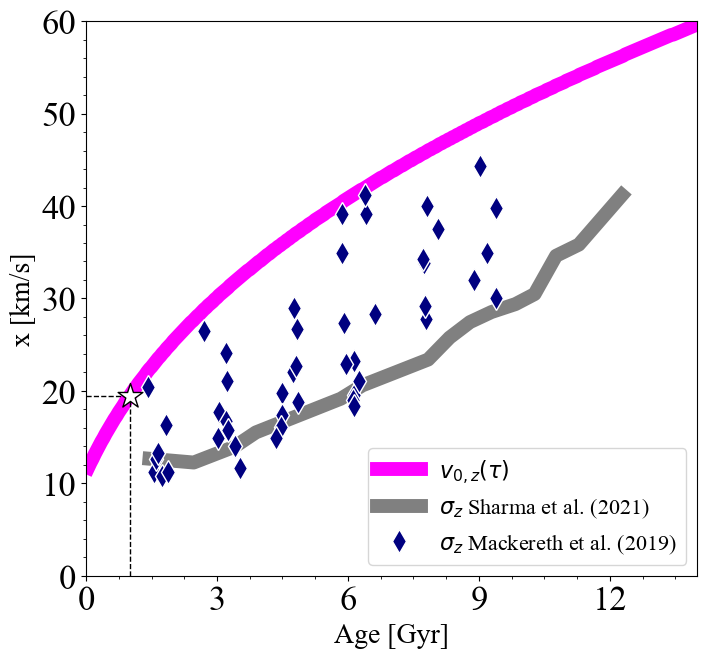}\\
\includegraphics[scale=0.40]{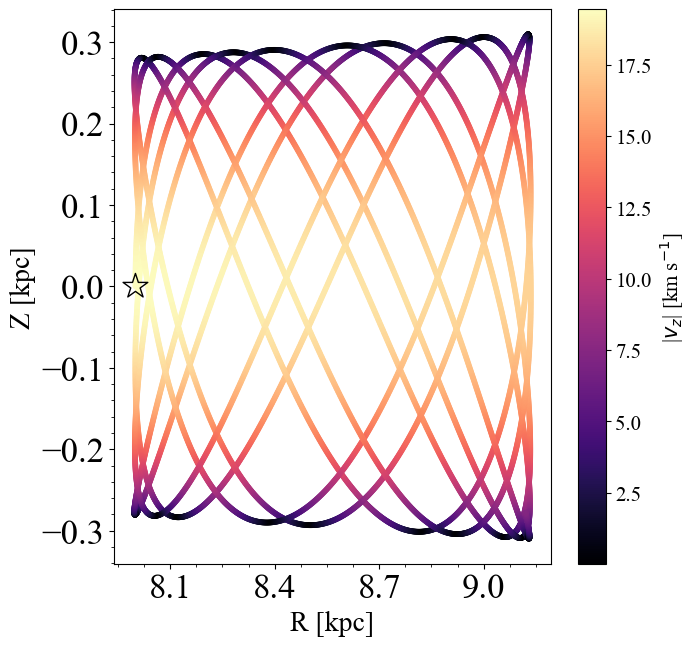}
\includegraphics[scale=0.38]{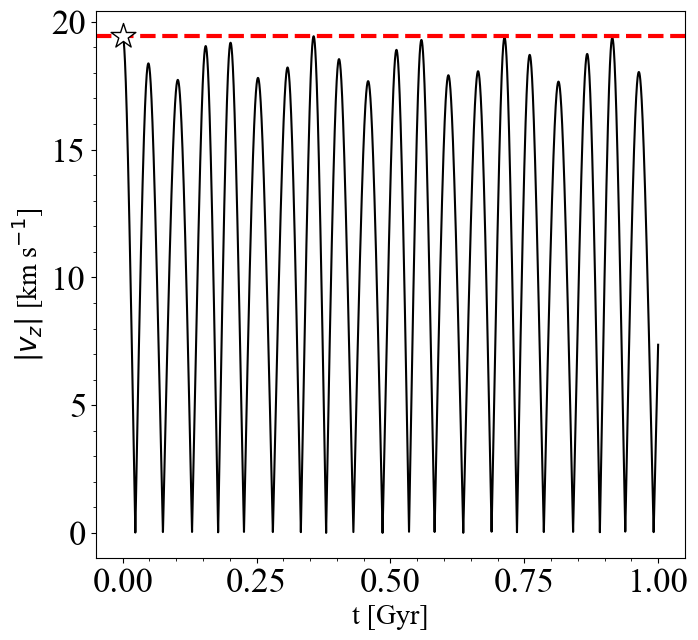}
\caption{{\it Upper panel:} The computed initial vertical velocity  $v_{0,z}(\tau)$ (which satisfies the condition of Eq. \ref{vo_ref}) as the function of the Galactic age $\tau$  for the SSPs born at different evolutionary times are reported with the solid magenta line.    Data for the age-velocity dispersion  $\sigma_z$  relation  in the solar vicinity from  \citet{sharma2021}  and  \citet{mackereth2019} are drawn with the grey line and diamond blue points, respectively.  {\it Lower left panel:} Orbit in the meridional plane $(R,Z)$ for the SSP  born in the Galactic plane ($z=0$ kpc) at 8 kpc (labelled with a star) at the Galactic time  $t_B=t_G-1$ Gyr and  integrated for 1 Gyr. The colour-coding stands for the modulus of the vertical velocity component $|v_z|$.  {\it Lower right panel:} The 1 Gyr evolution of the  modulus of the vertical velocity component $|v_z|$ for the same SSP as the lower left panel. The starred symbol   indicates the vertical velocity associated at the  birth  $v_{0,z} (\tau=1$ Gyr) (also indicated in the upper panel with the same symbol) which also corresponds to  the maximum value of the   vertical velocity component $|v_z|$ (see red dashed horizontal line).  
}

\label{vert_v}
\end{centering}
\end{figure*}
\begin{figure*}
\begin{centering}
\includegraphics[scale=0.49]{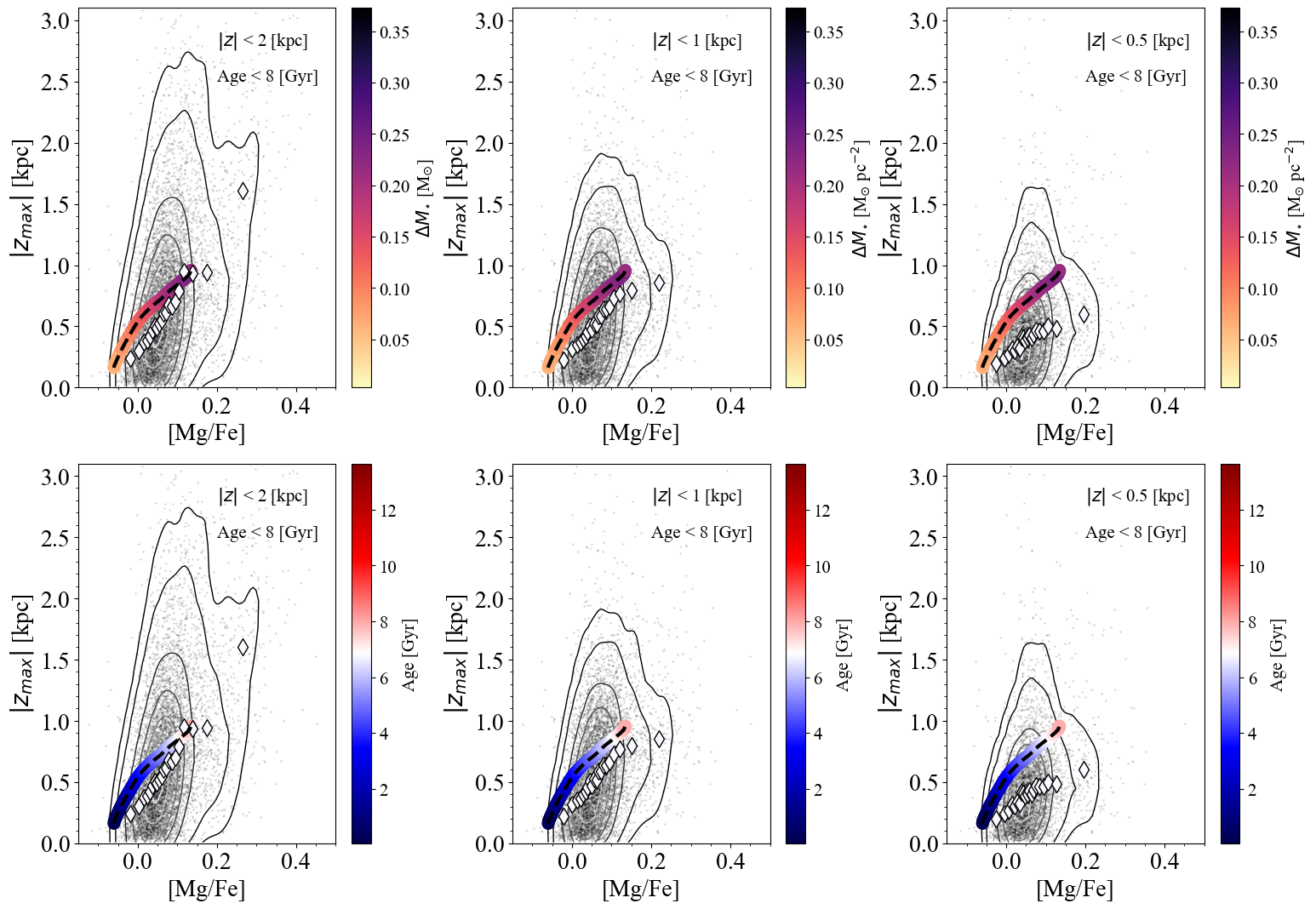}
\caption{
The grey points indicate stars with observed [Mg/Fe]  abundance ratios for stars from APOGEE DR16 \citep{Ahumada2019}  in the Galactocentric region between 6 and 10 kpc as a function of the computed maximum vertical heights \zmax with ages < 8 Gyr as reported in  the astroNN  catalogue. White diamonds stand for the median \zmax and [Mg/Fe] values in bins of [Mg/Fe] with the same  number of stars. The contour lines enclose fractions of
0.95, 0.90, 0.75, 0.60, 0.45, 0.30, 0.20 and 0.05
of the total number of observed  stars.
Stars with observed  vertical heights $|z|<2$ kpc, $|z|<1$ kpc and $|z|<0.5$ kpc,  are reported in the left, middle and right panels, respectively.   ES21 model predictions only for ages smaller than 8 Gyr including   the $J_z$ versus Age  relation  by \citet{ting}  is  indicated in each panel  with the dashed black line. 
In the upper panels the colour-coding indicates    the predicted surface stellar mass density $\Delta M_{\star} $ formed in age intervals of $0.05$ Gyr.  In lower panels, the colour coded circles depict    ages of new SSPs   formed during the Galactic evolution in age intervals of 0.05 Gyr. Although the  upper limit for the maximum vertical \zmax in all panels has been fixed at the value  3 kpc excluding  outlier stars, the median values and the contour density lines have been computed taking into account all the stars in the respective samples. 
}
\label{ES21_8}
\end{centering}
\end{figure*}

\subsection{On the conservation of the vertical action $J_z$ for Galactic disc stars}
\label{cons}
The existence of a non-classical integral of motion, associated with the actions of the
system, including the vertical action $J_z$, has been widely discussed by \citet{binney1984}.
Moreover, in presence of a  Galactic potential that changes slowly with time, the vertical action can be considered as an  adiabatic invariant.
For instance, 
the cylindrical adiabatic approximation  was introduced by
\citet{binney2010} to describe the distribution of stars in the Galactic
disc. The author argued that since the vertical frequency  of a disc star
is significantly larger than its radial frequency, the potential that affects vertical oscillations may be considered to vary slowly as
the star oscillates radially, with the consequence that $J_z$ is adiabatically invariant.
Hence, if  the vertical action $J_z$ is an adiabatic invariant, it  can also be  considered as an approximate invariant under radial migration through churning \citep{carlberg1987,sellwood2013}.
For this reason, it seems appropriate  to characterise the vertical motions of stars
by their vertical actions, $J_z$, rather than their vertical velocities
$v_z$ or velocity dispersion $\sigma_z$ (in fact, $\sigma_z$ will
change in a growing potential and under radial migration).
 
 \begin{figure*}
\begin{centering}
\includegraphics[scale=0.49]{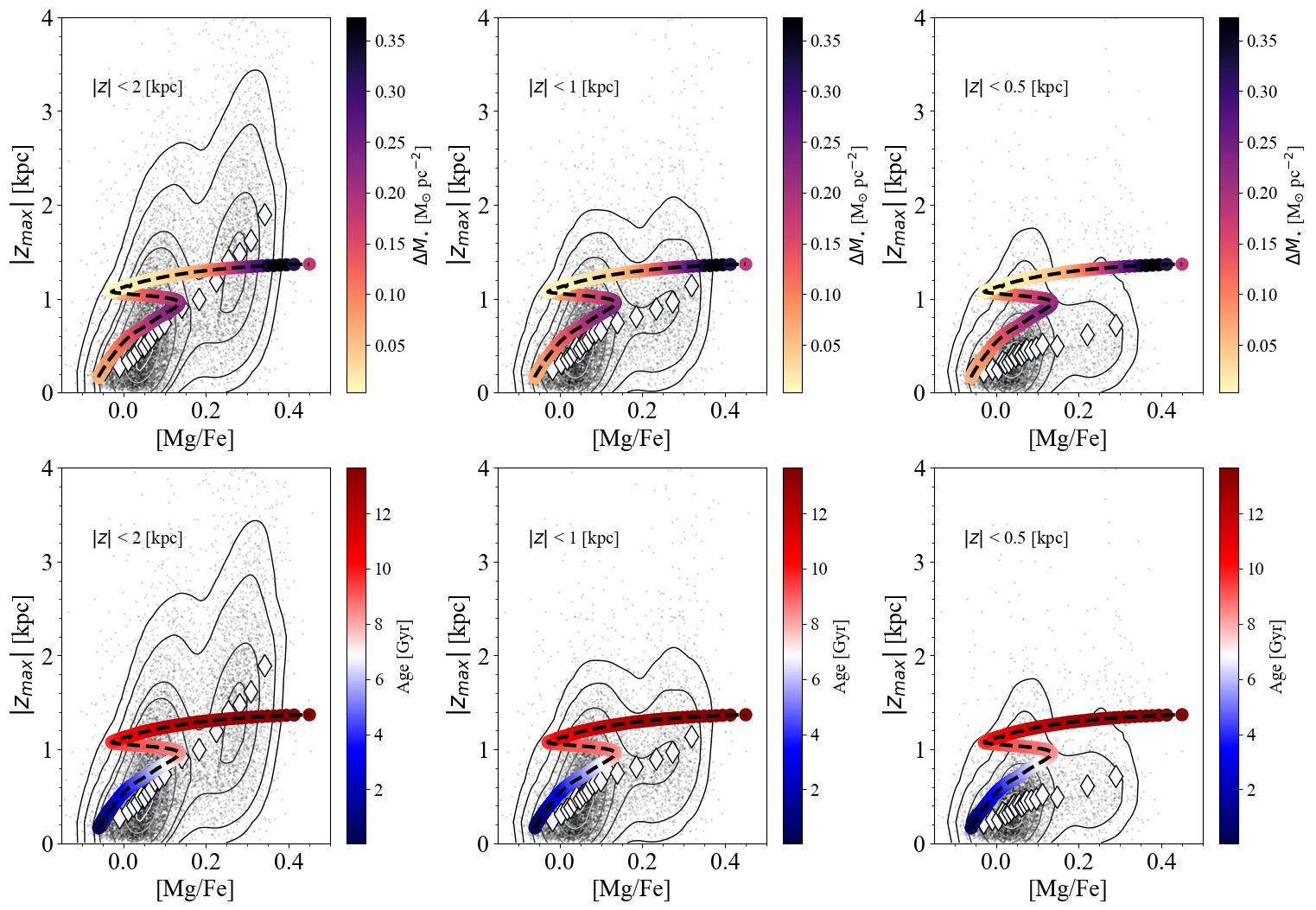}
\caption{ Same as  Fig. \ref{ES21_8} but   also for  observed and predicted stars older  than 8 Gyr have been depicted extending the \citet{ting} relation to the whole stellar ages (see Fig. \ref{jz_zmax}).
In the figure, the maximum vertical height is shown between 0 and 4 kpc for the sake of clarity, however the median values and the contour density lines have been computed using all stars (including those with \zmax  greater than 4 kpc).
 }
\label{ES21all}
\end{centering}
\end{figure*}
 
In principle, external  spiral galaxies characterised by high frequency of   vertical motion compared to in-plane evolution should be the perfect candidates  where 
vertical action $J_z$ is a 
conserved quantity. However,  \cite{Solway2012}  showed that for
isolated galaxies the vertical action is not a constant of motion of
individual stars, and is only conserved on average for a sample of
stars, with an intrinsic dispersion of $\sim 20\%$.
Furthermore, \citet{vera2016}  using high-resolution N-body simulations of  Milky Way-like discs showed that stars deviate from near-circular orbits, reducing the degree at which the actions are conserved for individual stars.

Bearing in mind all the above-mentioned caveats  for late-type galactic systems in presence of spiral arms like the Milky Way, in the next Section we will compute orbits for Galactic disc SSPs born at different times considering  the vertical action  as  a conserved quantity for coeval stars.

\subsection{Computing the orbital maximum  vertical  excursion and the vertical [Mg/Fe] gradient}
\label{zmax_c}

 We assume for the Galaxy the three component steady-state {\tt MWPotential2014}  gravitational potential  of \citet{bovy2015}.
 The potential model consists of: i) a bulge modeled as a power-law density profile that is exponentially cut-off; ii) a  Miyamoto-Nagai \citep{miyamoto1975}  disc; and iii)  a dark-matter halo described by the Navarro, Frenk \& White (NFW, \citealt{navarro1996})  profile. All parameters and properties of {\tt MWPotential2014}  potential are listed in  the Table 1 of \citet{bovy2015}.  

Following the discussion in Section \ref{cons},  we impose that  each SSP born at 8 kpc in the Galactic plane and initial vertical height coordinate $z=0$ kpc   at a certain Galactic evolutionary time $t_B$,  conserves the average vertical action  $\widehat{J_z}$ in their orbit 
 subject to the gravitational potential of the Galaxy.
 
  In the calculations, we consider SSPs formed in constant age intervals $\Delta \tau$ fixed at the value of 0.005 Gyr (the   total  number of SSPs is 2740), which is  identical to the time-step of the chemical evolution model. 
Recalling the \citet{ting} relation introduced  in Section \ref{tings} and associated discussion, the value of conserved vertical action $\widehat{J_z}$   for a SSP with age $\tau=t_G-t_B$  can be estimated using Eq. (\ref{eq_ting}), 
 where  $t_G$  is the age of the Galaxy. 

 We integrate stellar orbits for different SSPs($\tau$) using the  \textsc{galpy}\footnote{http://github.com/jobovy/galpy}  package subject to  the  {\tt MWPotential2014}  gravitational potential    \citep{bovy2015}.
 For the rotational velocity $v_T$ at the solar distance $R_{\odot}$ we use the one computed by  \citet{ablimit2020} applying the 
  three-dimensional velocity vector  method for Cepheids ($v_T=232.5$ ± 0.83 $\mathrm{km}\,{{\rm{s}}}^{-1}$ ) that  is consistent with the most recent estimation by \citet{nit2021}  with a dynamical model of the Milky Way using APOGEE and Gaia data. 
The initial vertical velocity for the SSP $v_{0, z}(\tau)$ born at the evolutionary time $t_B=t_G-\tau$ has been chosen by satisfying the following condition:

\begin{equation}
  \Bigl|\quad\underbrace{J_{z}\Bigl(R_{\odot}, v_{0,z}(\tau) \Bigr)}_{\mbox{\textsc{galpy}}} \qquad-   \underbrace{\widehat{J_z}(R_{\odot}, \tau)}_{\mbox{\citet{ting}}} \Bigr|<0.05  \, \mbox{    [kpc km s}^{-1}].
  \label{vo_ref}
\end{equation}
i.e.,  the computed vertical action $J_z$ with  \textsc{galpy}  and the one  with Eq. (\ref{eq_ting})  differ less than  $0.05$   [kpc km s$^{-1}$].
In Fig. \ref{jz_zmax}, we show the maximum  vertical  excursion  from  the  midplane \zmax as a function of the SSP ages and vertical action with an age resolution of $\Delta \tau=0.05$ Gyr assuming that initial vertical height is $z=0$ kpc at any Galactic time (all the stars are born exactly right in the disc plane).

One of the most important conclusions of \citet{ting} is that orbital scattering is a plausible and viable mechanism to explain the age-dependent vertical motions of disc stars. Hence, here we assume that the star formation  happens  in the disc plane with null vertical height ($z=0$ kpc) where the infalling gas gets accreted (see Eq. \ref{a}). Higher vertical coordinates \zmax (see Fig. \ref{jz_zmax}) can be reached only  by older SSPs   because of the kinematic scattering. Here, we are ignoring the extra-heating processes from merging events in which stars from galactic systems are engulfed by the Galaxy \citep{helmi2018}.

In Fig. \ref{vert_v}, we show the initial vertical velocities  $v_{0,z}(\tau)$  as the function of the Galactic age for the SSPs born at different evolutionary times $t_B= t_G-\tau$. In presence of the steady-state gravitational potential  {\tt MWPotential2014}, the initial vertical velocity of one SSP   formed in the Galactic plane (with initial vertical height $z=0$ kpc) in the solar neighborhood also corresponds  to the maximum vertical velocity  of the  orbit.
 In fact, in Fig. \ref{vert_v} we also show  the temporal evolution of the modulus of the vertical velocity component of  the SSP  born in the Galactic  plane ($z=0$ kpc) at 8 kpc at the Galactic time  $t_B=t_G-1$ Gyr and  integrated for 1 Gyr. It is evident that  the vertical velocity at the SSP birth $v_{0,z} (\tau=1$ Gyr) corresponds to  the maximum value of the   modulus of vertical velocity  $|v_z|$.

Because of the symmetry of the system, we can  consider this quantity as an estimate for the  upper limit of  the   vertical velocity dispersion $\sigma_z$  at the Galactic age $\tau$. Hence, considering all the SSPs born at different evolutionary times we can estimate  the upper limit of the observed  vertical velocity dispersion $\sigma_z$ versus  Galactic age $\tau$ relation. In Fig. \ref{vert_v}, we compare  the predicted $v_{0,z}(\tau)$ values with the  age-$\sigma_z$ relations  observed in the solar vicinity by \citet{mackereth2019} analysing APOGEE stars  and by \citet{sharma2021} for GALAH objects. 
In Fig. \ref{vert_v}, each  \citet{mackereth2019}  data point represents a mono-[Fe/H] bins that have more than 200 stars.
We note that the computed initial velocity distribution traces very well  the upper limit of the age-$\sigma_z$ relation.

  As  mentioned in the Introduction, the ES21 model predicts that the  Galactic disc follows  an inside-out formation and consequently  the Galactic potential should evolve as a function of time as well. 
 In the Milky Way-like galaxy  in the cosmological context presented by  \citet{bird2013} and characterised by an inside-out formation,    the surface stellar mass density and the median height above the disc (of stellar populations born at different Galactic ages)  did not vary much in the last $\sim$ 8 Gyr at the solar distance. This  implies  that the inside-out formation did not affect substantially  the Galactic potential at Galactic ages < 8 Gyr. Hence,  a static potential for the integration of the orbit of stars with ages < 8 Gyr (see Fig. \ref{ES21_8}) is a valid approximation. However,  this could not be true when we extend the \citet{ting} relation to older ages, then it is important to stress this caveat of our approach.

 In the next Section we will show model predictions  combining together the 
  [Mg/Fe] abundance ratio for stars born at different Galactic times (i.e. with different ages $\tau$) as predicted with chemical evolution models presented in Section \ref{s:chemmod} with the maximum orbital height above the plane \zmax($\tau)$ computed as described in Section \ref{zmax_c}.

\begin{figure}
\begin{centering}
\includegraphics[scale=0.46]{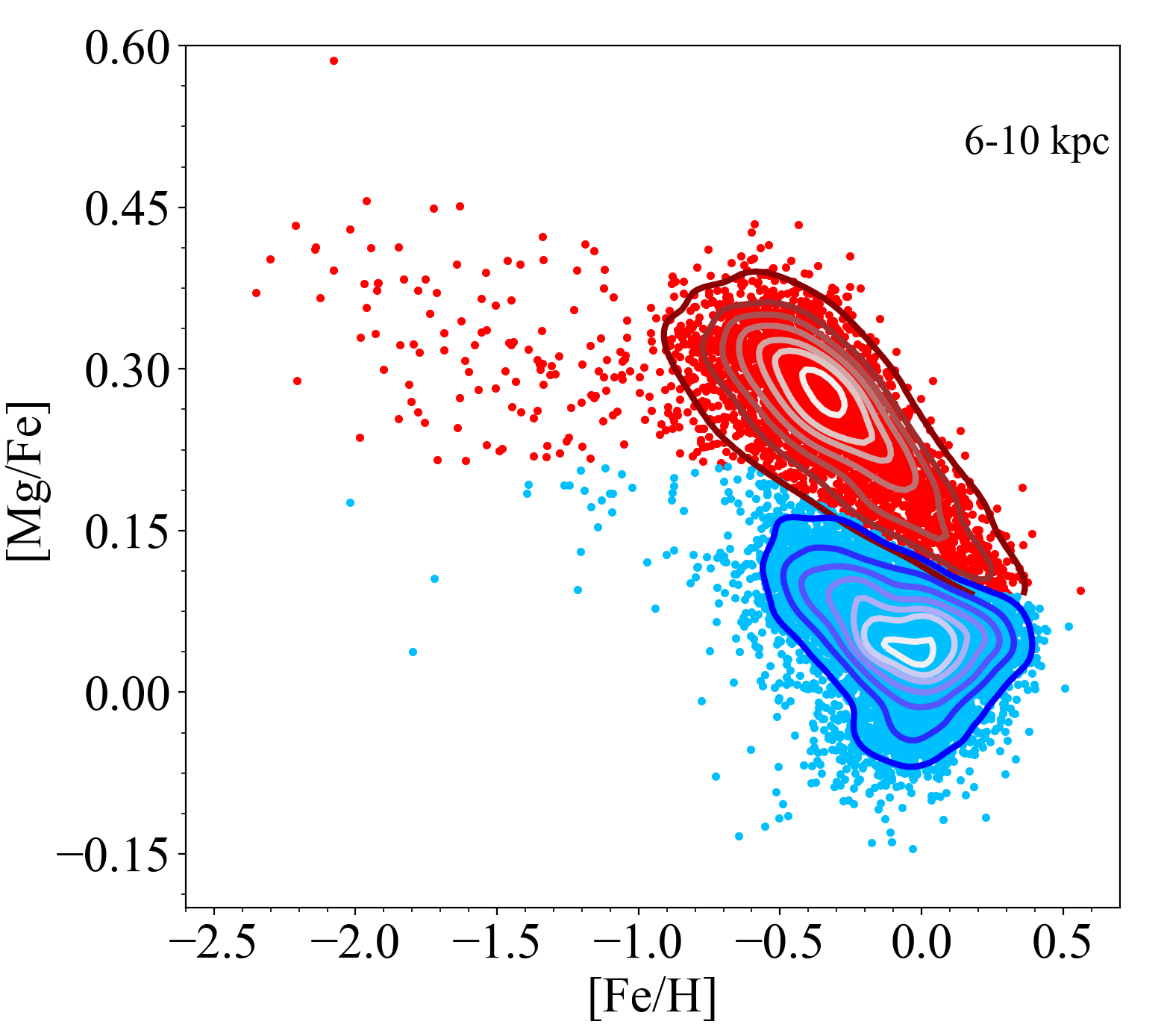}
\caption{  Disc components selected based on chemistry of the APOGEE stars in the Galactic region enclosed between 6 and 10 kpc presented in Section \ref{data:s}.  With the red points are shown high-$\alpha$ sequence stars, whereas with light blue the low-$\alpha$ ones. The contour lines enclose fractions of 0.95, 0.90, 0.75, 0.60, 0.45, 0.30, 0.20 and 0.05 of the total number of observed stars for the two sequences, separately. 
}
\label{sele}
\end{centering}
\end{figure}

\begin{figure*}
\begin{centering}
\includegraphics[scale=0.42]{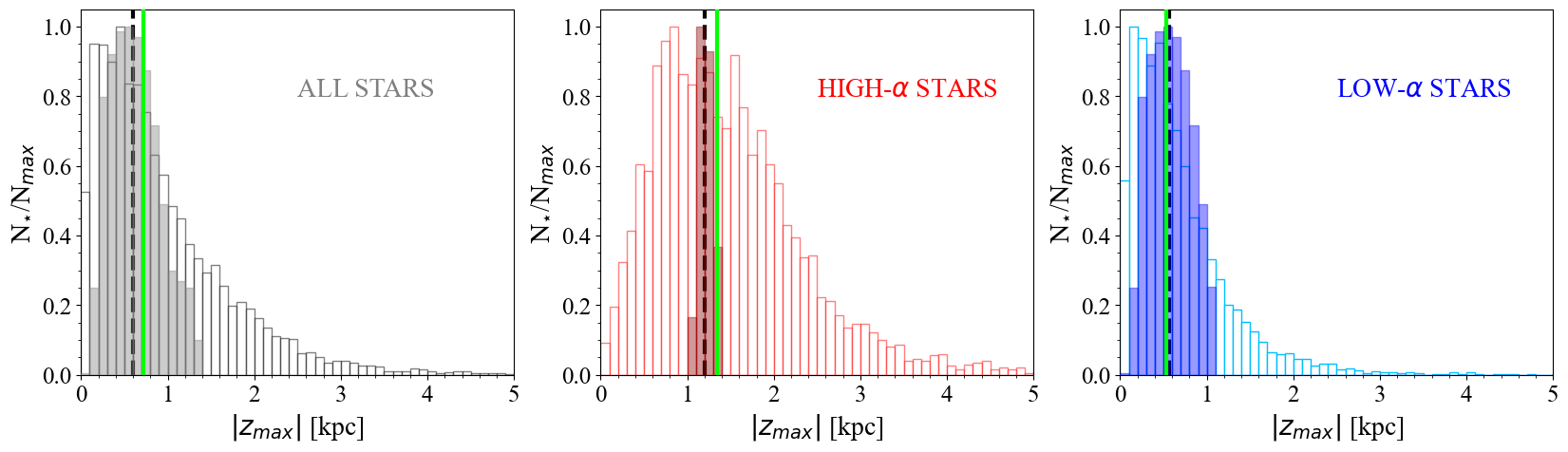}
\includegraphics[scale=0.42]{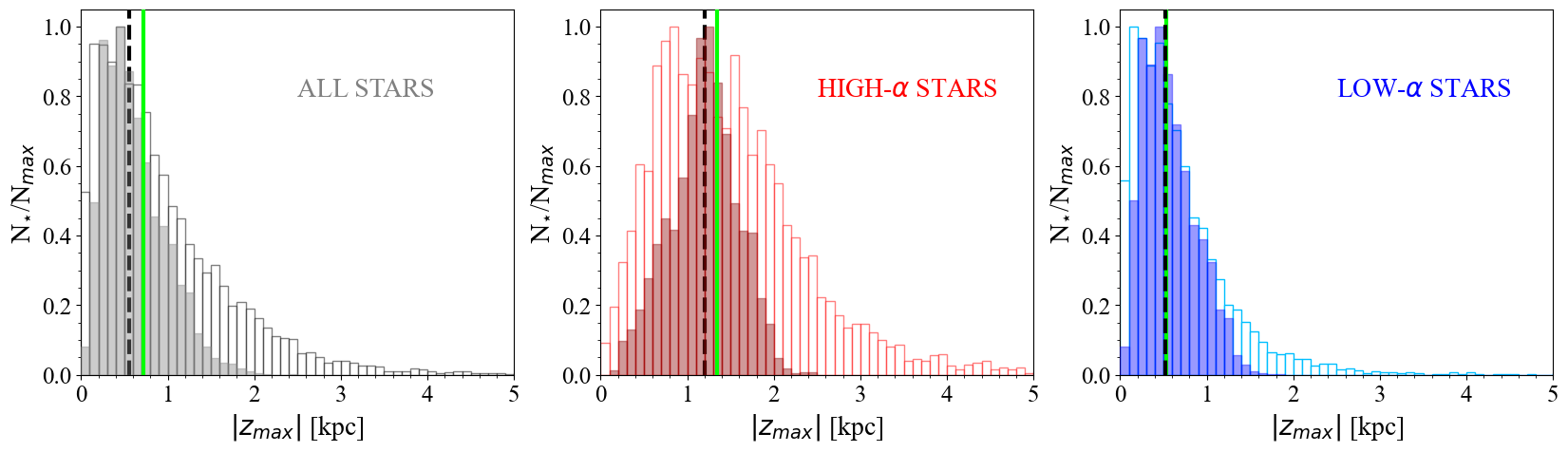}
\includegraphics[scale=0.42]{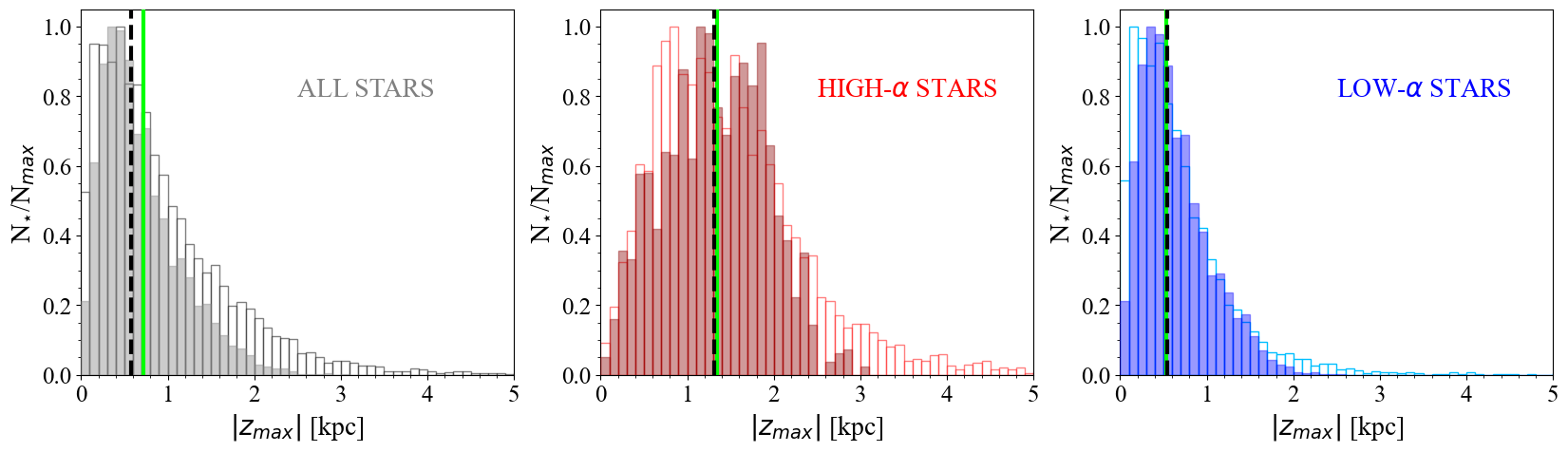}
\caption{ Comparison between the    predicted  $|z_{max}|$  distributions by the ES21 model (filled histograms) with APOGEE+astroNN stars (empty histograms) for   the whole stellar sample  introduced in Section \ref{data:s} (left panels), high-$\alpha$  sequence (middle panels), low -$\alpha$ sequence (right panels) normalised to the maximum number of stars in the bins of the distributions. In the first row we report ES21 predictions  using  the vertical action $J_z$ reported in Eq. \ref{eq_ting} without including any errors. In the second row, we report ES21 model results adding  a random error normally distributed around $J_z$ with  standard deviation $\sigma_{J_z}$ fixed at the values of  $\sigma_{J_z}=0.5 \cdot J_z$, and finally in the last row the case with $\sigma_{J_z}= 1 \cdot J_z$ is drawn. In each panel, the vertical  solid green and dashed black lines indicate the median values of APOGEE stars and model predictions, respectively. 
  }
\label{ES21_distrib}
\end{centering}
\end{figure*}

\begin{figure}
\begin{centering}
\includegraphics[scale=0.4]{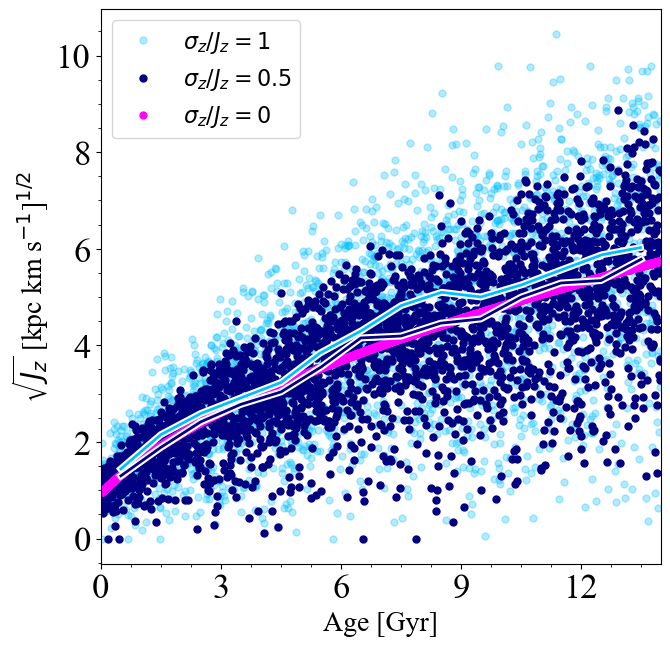}
\includegraphics[scale=0.4]{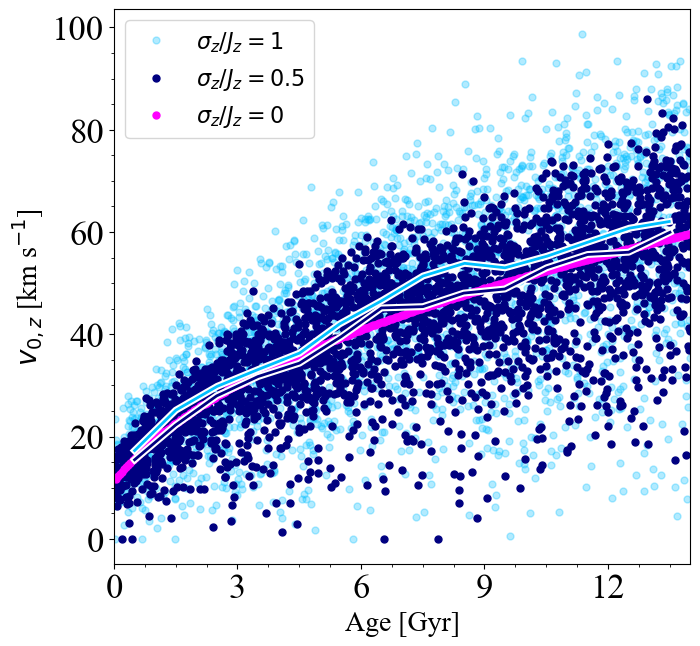}
\caption{ {\it Upper panel:}  The \citet{ting} relation between the vertical action $J_z$  and the stellar age,  computed at 8 kpc is indicated with the magenta line. The   ``new'' vertical action  $\widehat{J}_{z, \mbox{ new}}$  of eq. (\ref{sigma_J})  including observed dispersion for   $\sigma_{J_z}/J_z=0.5$ and   $\sigma_{J_z}/J_z=1$ cases are reported with the dark and light-blue points, respectively.
The solid dark-blue (light-blue) line indicates the medians values for $\sigma_{J_z}/J_z=0.5$ ($\sigma_{J_z}/J_z=1$).
{\it Lower panel:} The computed initial vertical velocity  $v_{0,z}(\tau)$ (which satisfies the condition of Eq. \ref{vo_ref}) as the function of the Galactic age $\tau$  for the SSPs born at different evolutionary times are reported with the solid magenta line. As the upper panel, the computed $v_{0,z}$  including observed dispersion for   $\sigma_{J_z}/J_z=0.5$ and   $\sigma_{J_z}/J_z=1$ cases are indicated with the dark and light blue points, respectively.
}
\label{jz_age_disper}
\end{centering}
\end{figure}

\begin{figure*}
\begin{centering}
\includegraphics[scale=0.35]{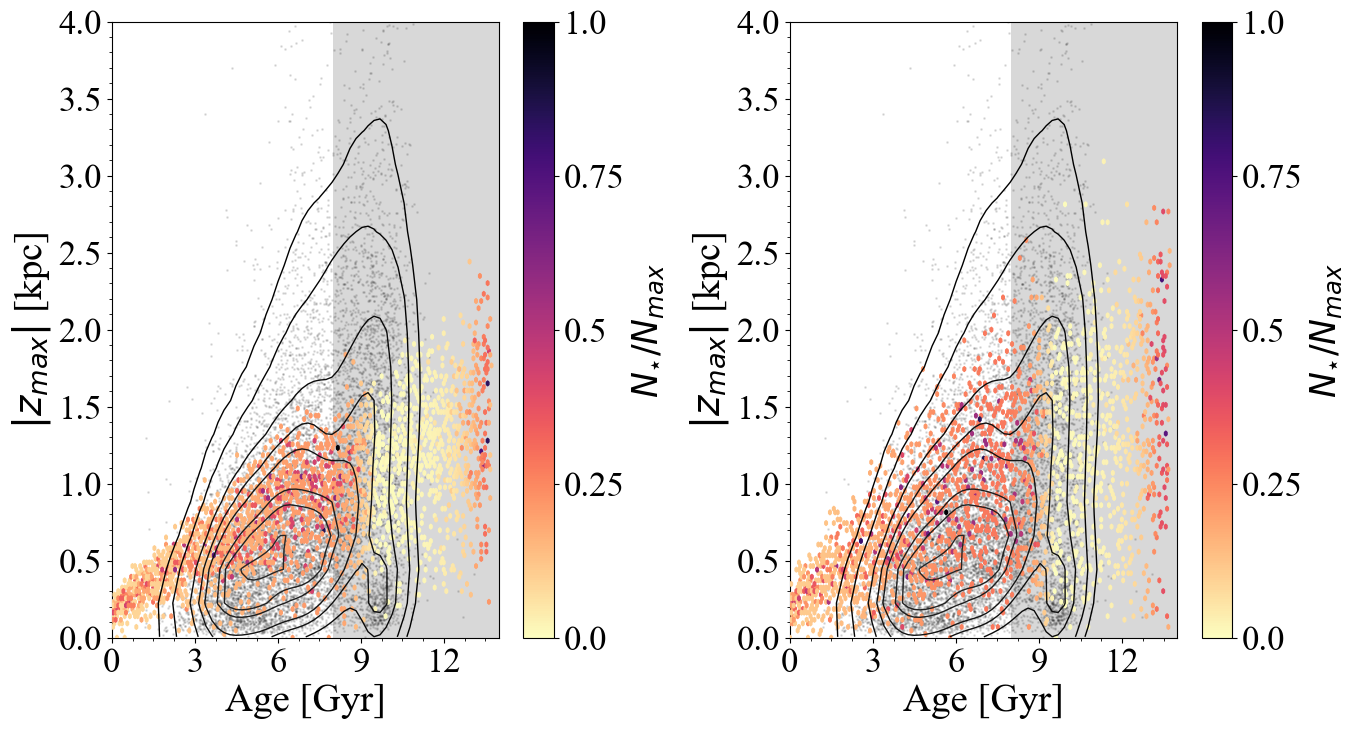}
\includegraphics[scale=0.35]{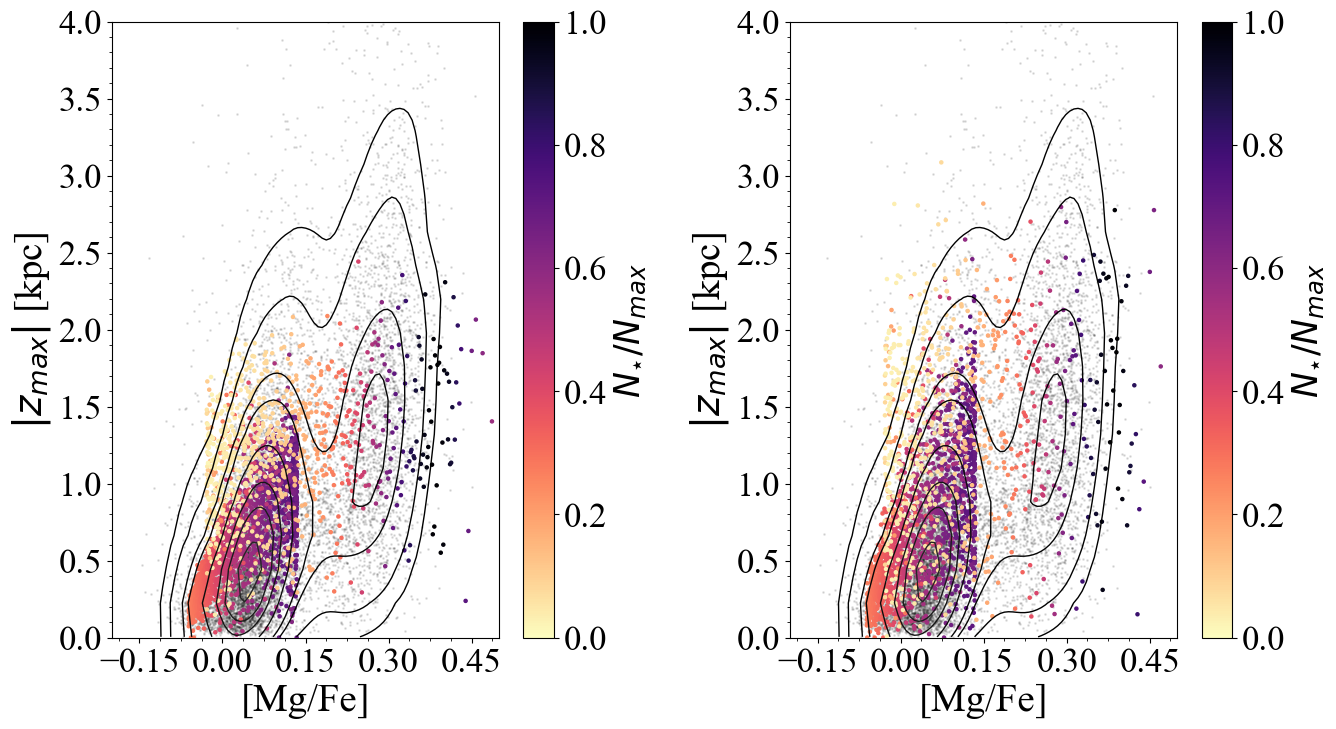}
\caption{  {\it Upper panels:}} The vertical maximum excursion  $|z_{max}|$  versus age relation for the ES21 model. The grey points indicate stars within Galactocentric region between 6 and 10 kpc and  $|z|<2$ kpc  as reported in the astroNN catalogue. The contour lines enclose fractions of 0.95, 0.90, 0.75, 0.60, 0.45, 0.30, 0.20 and 0.05 of the total number of observed stars.  The colour coding represents the total number of stars predicted by the ES21 model in that region. In the left panel we draw model results  considering a Gaussian error (see Eq. \ref{sigma_J}) with standard deviation  $\sigma_{J_z}= 0.5 \cdot J_z$, whereas in the right one we assume $\sigma_{J_z}= 1 \cdot J_z$.  The shaded grey  area highlights the region with ages larger than 8 Gyr, for which the use of Eq. (\ref{eq_ting}) may not be justified.  {\it Lower panels:} As the upper ones but for the $|z_{max}|$  versus [Mg/Fe] relations.  
\label{ageES21}
\end{centering}
\end{figure*}

\section{Results}
\label{s:results} 
We compare the  vertical [Mg/Fe] abundance distribution in the solar vicinity predicted by  chemical evolution models presented in  Section \ref{s:chemmod} to the observed APOGEE DR16 stellar abundance ratios and the associated orbital properties and ages as presented in the  astroNN catalogue (see Section \ref{data:s}). 
We show results for   both in the case where all stars are younger than 8 Gyr and in the case including all ages.  We recall that \citet{ting} relation for the vertical action (Eq. \ref{eq_ting}) has been recovered for stellar ages smaller than 8 Gyr (thin disc stars). Here, we  test  whether extending this  relation  to older stellar populations allow us to reproduce the vertical distribution of the [Mg/Fe] abundances in  APOGEE data for the full age range.
\subsection{ES21 model}
\label{s:ES21}
In Fig. \ref{ES21_8}, we show   vertical abundance distributions   using     ES21 model computed at 8 kpc   as expressed by  relation between the maximum vertical  height \zmax (see the methodology in  Section \ref{zmax_c}) versus the [Mg/Fe] abundance ratios for stars with ages < 8 Gyr.  
We considered  different cuts in the observed Galactic vertical heights, i.e. $|z|<2$ kpc, $|z|<1$ kpc and $|z|<0.5$ kpc,  in order to better analyse the vertical structure.

Observed stars with astroNN ages younger than 8 Gyr are more likely  part of the low-$\alpha$ sequence. In fact, the vertical [Mg/Fe] abundance gradient seems to be  traced by  a single stellar sequence as visible from the contour density lines:  the contour which encloses  the 75\% of the stars presents the   absence of any elongated structure towards higher [Mg/Fe] values and  hence without trace of bimodality. 

In Fig. \ref{ES21_8}, we note that the observed  \zmax  versus  [Mg/Fe] relations  traced by   stars  having observed vertical heights  $|z|<2$ (left panels) and  $|z|<1$ kpc (middle panels)  are  well reproduced by our model predictions.
In contrast, smaller  \zmax values  than the computed one emerge from the data in the right panel with  $|z|<0.5$ kpc: in fact in this case   a consistent number of observed  low-$\alpha$ stars have been discarded. 

In our model computed in the solar neighborhood,  an age smaller than 8 Gyr corresponds to  an evolutionary time larger than 5.7 Gyr. In ES21, the quantity $t_{\rm max}$ (i.e. the delay between the peaks of the two gas infall) is $4.085$ Gyr (see Table \ref{tab_mcmc}). Hence,  with considering  only stars with  ages < 8 Gyr, we exclude   some thin disc stars, specifically the ones  formed as soon after  the beginning of the second gas infall (born at evolutionary time enclosed between 4.085 and 5.7 Gyr), and  totally ignoring the thick disc component.

In Fig. \ref{ES21all}, we show model results extending the \citet{ting} relation to high-$\alpha$ sequence stars (ages > 8 Gyr, i.e.    the grey shaded area in Fig. \ref{jz_zmax}).
The  stellar sample with the full age range of the astroNN catalogue     shows  a clear bimodality in the relation \zmax versus [Mg/Fe], more evident  with the cut $|z|<2$ kpc   (left panels in Fig. \ref{ES21all}).
Our model is able to predict such a dichotomy  displaying two different sequences in the \zmax versus [Mg/Fe]  relation.  
In the phase when the star formation resumes  immediately after the beginning of the second  gas infall,  ejecta from Type II SNe   produce a steep rise in the [$\alpha$/Fe] ratio, followed by a decrease  due to pollution from Type Ia SNe  as visible in  Fig. \ref{ES21all}. This kind of transition between the two disc sequences is not obvious in   the data. However, as pointed out by the coloured coded points indicating the  stellar mass $\Delta M_{\star}$ produced in constant age interval of $\Delta \tau=0.05$ Gyr, the "upturning" feature in the chemical evolution track is characterised by  a low stellar mass  content compared to  other   Galactic evolutionary phases.

In Fig. \ref{ES21all}, we see that   the predicted curve for the  vertical  [Mg/Fe] distribution  by the model nicely overlaps with the regions with the highest densities of stars in both low-$\alpha$ and high-$\alpha$ APOGEE stars with observed heights $|z|<2$ kpc  (left panels).
However, the observed median \zmax  values  (computed in bins of [Mg/Fe] with the same number of stars) for high-$\alpha$ stars are larger than our model results. In fact,  we predict a too flat (almost constant) growth of the maximum vertical height \zmax for  thick disc sequence stars.

One possible explanation for such a discrepancy is that the stellar sample could be contaminated by the presence of halo stars. However, it is more likely  that   other dynamical processes - such as merger and accretion episodes - have played a crucial role in perturbing stellar orbits of the high-$\alpha$ sequence in addition to simple scattering processes   as assumed by the \citet{ting} model. for thin disc stars (stellar ages < 8 Gyr).
Cosmological simulations have shown that Milky Way-like galaxies  are frequently experiencing minor mergers 
\citep{quinn1993,walker1996, velazquez1999, kaza2009, house2011,gomez2013,donghia2016, Moetazedian2016}, external perturbations that can heat up the disc.  
\citet{helmi2018} demonstrated that the inner halo is dominated by debris from an accreted  object more massive than the Small Magellanic Cloud, alias Gaia-Enceladus  \citep{vincenzo2019}. Gaia–Enceladus must have led to the dynamical heating of the precursor of the Galactic thick disc (stars with the same [$\alpha$/Fe] versus [Fe/H] as thick disc  but with  different kinematics).
Moreover, radial stellar migration
\citep{sellwood2002, schoenrich2009MNRAS, minchev2010}
 should cause more extended vertical motion at
reduced velocity dispersion for stars that move outward, and the opposite effect for stars that move inward 
\citep{loebmn2011,minchev2012}.

In Fig. \ref{ES21all}, we note that the median values  of the observed vertical [Mg/Fe] distribution for  $|z|<1$ kpc (middle panels) and $|z|<0.5$ kpc case (right panels) show  an evident change in the slope  as we moved from the low-$\alpha$ (steeper) to high-$\alpha$ stars (flatter) more in agreement with our model predictions compared with $|z|<2$ kpc cut (left panels). Hence, it is likely that applying  $|z|<1$ kpc  and $|z|<0.5$ kpc cuts, we are excluding from the APOGEE sample  objects that possibly have been affected by  extra-heating from  past merging  events.

We conclude that, if we consider as high-$\alpha$ sequence stars  the objects with $|z|<2$ kpc,
 some  extra-heating  from  gravitational perturbers (i.e., the constellation of clusters, small dwarf galaxies)   should be taken into account to achieve a better agreement between model results and  observed \zmax versus [Mg/Fe] median values.  Recently,  \citet{conroy2021} showed that the Large Magellanic Cloud (LMC)   had an impact on the position of the centre of mass of the Milky Way + LMC system, and  creating important  dynamical signatures in the MW (and LMC) halo as well.

 In order to further investigate the validity of the proposed two-infall model, we  compare the observed and predicted $|z_{max}|$ distributions  (studying only the case with $|z|<2$ kpc because of the evident bimodality in the [Mg/Fe] versus $|z_{max}|$ space) assuming a  separation between  high-$\alpha$ and low-$\alpha$ stars based on the chemistry.  
In Fig. \ref{sele},  we report the  APOGEE-DR16 stars with the same selection criteria as presented in Section \ref{data:s}, where we impose a disc dissection based on chemistry similar to the one suggested by \citet{victor2018}  for the APOKASC sample and by \citet{ness2019}. 

In the first row of  Fig. \ref{ES21_distrib},  we  compare  $|z_{max}|$ distributions   predicted by the ES21 model  with  data   for the whole stellar sample (left panel),  high-$\alpha$ (middle panel) and low-$\alpha$ sequence stars (right panel), respectively. Following the same procedure adopted in \citet{spitoni2019}, we assume that   SSPs formed at evolutionary times smaller (larger) than $t_{max}$ (time delay between the two gas infall events) are considered as part of the high (low)-$\alpha$ sequence.   As in Fig. \ref{ES21all} for the \zmax versus [Mg/Fe] distributions, here  we are primarily interested in  reproducing the trend of the observations. 
In fact, in Fig. \ref{ES21_distrib} the predicted median values by the ES21 model are consistent with the APOGEE DR16+astroNN catalogue stars.
However, we notice that especially for the  high-$\alpha$ case  we fail in reproducing the large spread of the observed distribution.

\begin{figure*}
\begin{centering}
\includegraphics[scale=0.49]{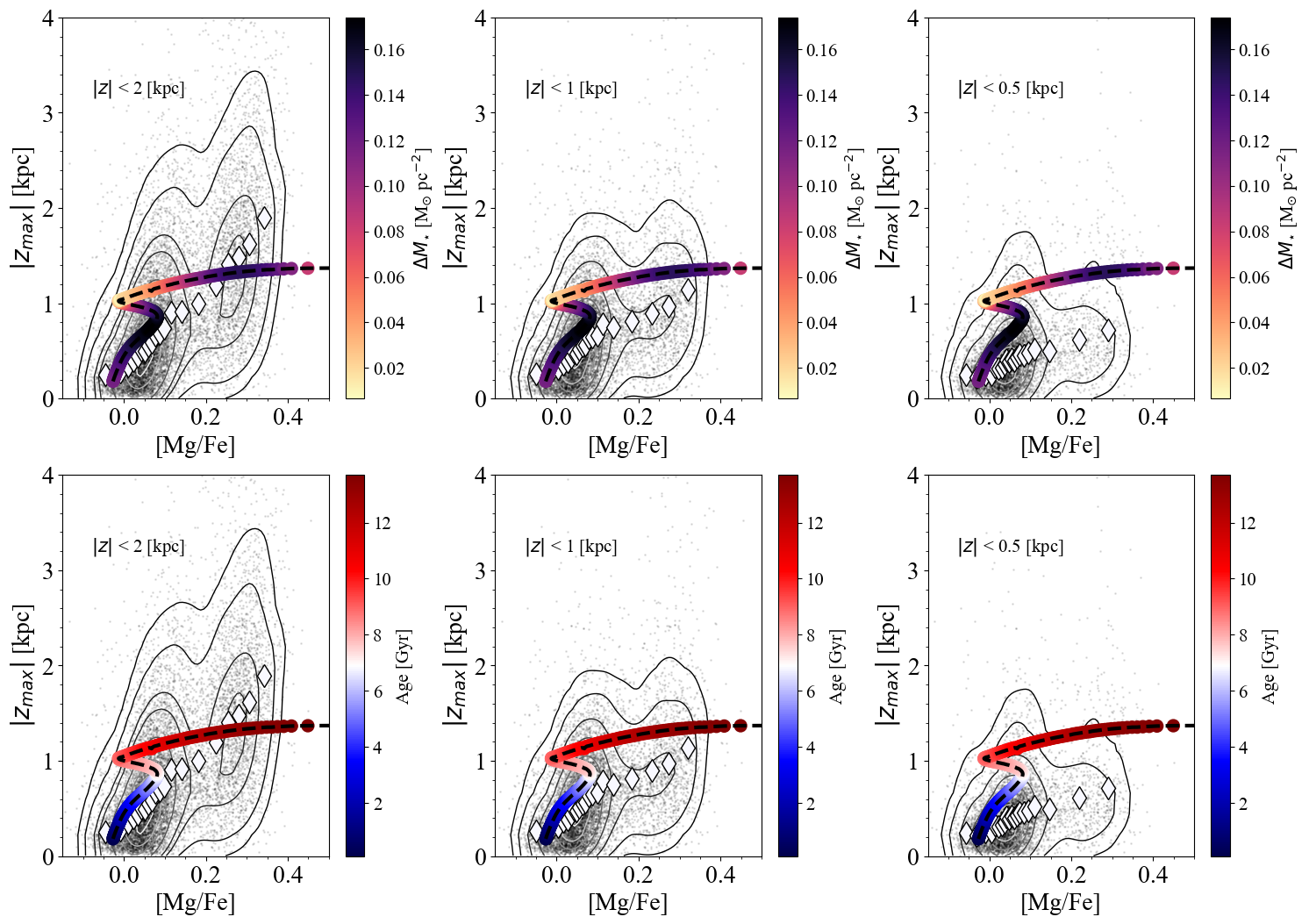}
\caption{ As in Fig. \ref{ES21all} but  considering chemical evolution results of the model \emph{M2} of ES20.}
\label{ES20all}
\end{centering}
\end{figure*}
\begin{figure}
\begin{centering}
\includegraphics[scale=0.38]{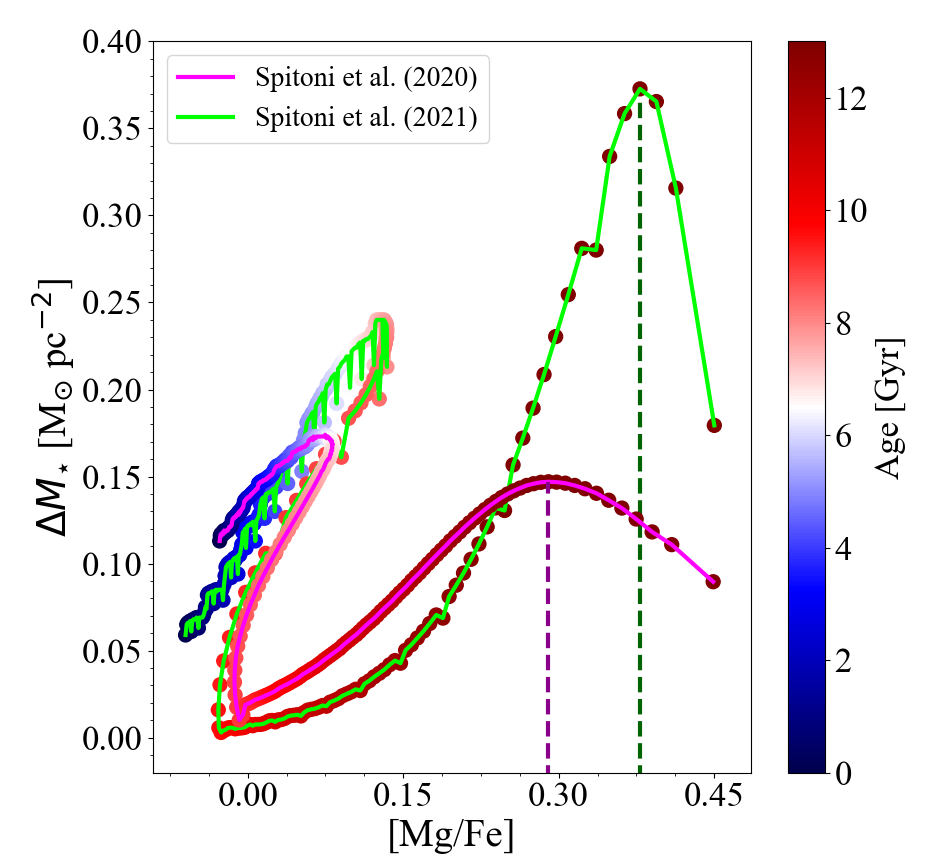}
\caption{Predicted surface stellar mass density $\Delta M_{\star}$ formed in age intervals of 0.05 Gyr as function of the [Mg/Fe] abundance ratio predicted by  ES21 model (green solid line) and by \emph{M2} model of  ES20  (magenta solid line). Colour-coded points correspond to different ages  of the SSPs. Vertical dashed lines depict the associated maximum $\Delta M_{\star}$ values  in the high-$\alpha$ sequence stars.   } 
\label{dm}
\end{centering}
\end{figure}

\begin{figure}
\begin{centering}
\includegraphics[scale=0.27]{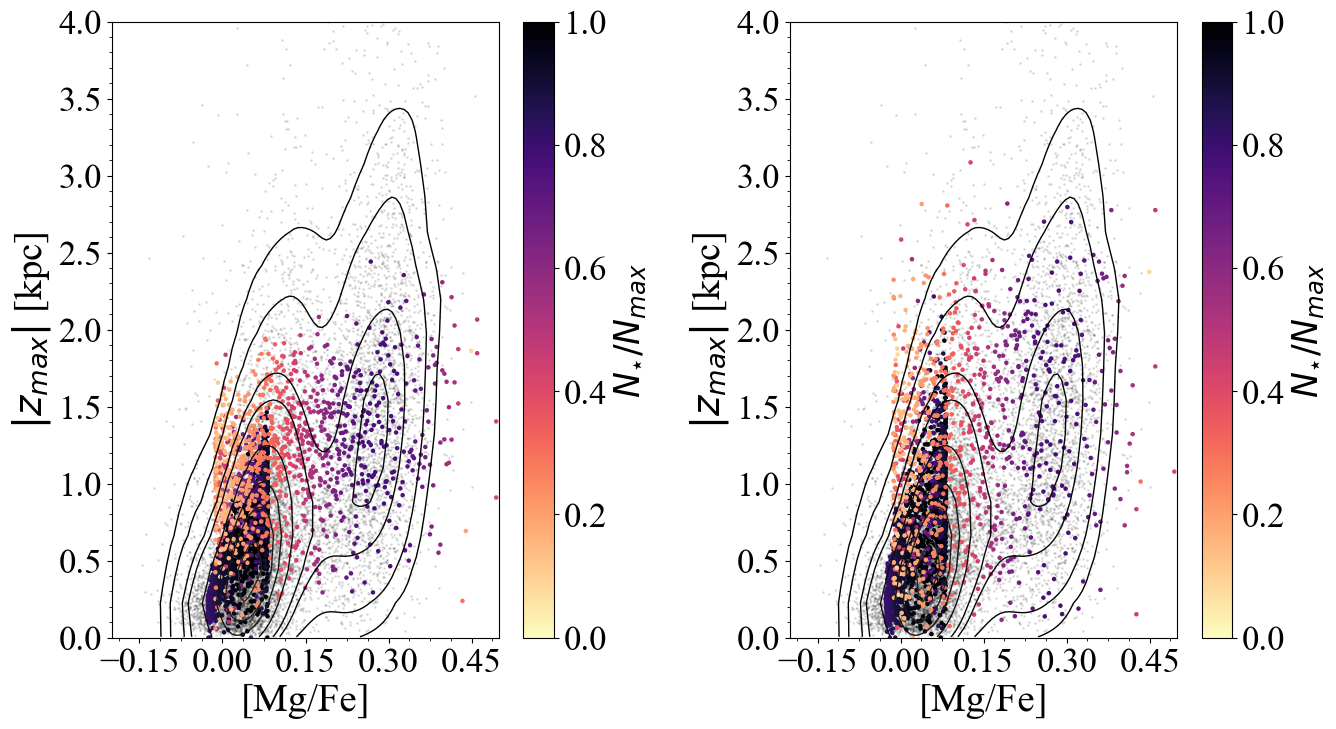}
\caption{  The vertical maximum excursion  $|z_{max}|$  versus [Mg/Fe] relation for the ES20 model. The grey points indicate stars within Galactocentric region between 6 and 10 kpc and  $|z|<2$ kpc  as reported in the astroNN catalogue. The contour lines enclose fractions of 0.95, 0.90, 0.75, 0.60, 0.45, 0.30, 0.20 and 0.05 of the total number of observed stars.  The colour coding represents the total number of stars predicted by the ES21 model in that region. In the left panel we draw model results  considering a Gaussian error (see Eq. \ref{sigma_J}) with standard deviation  $\sigma_{J_z}= 0.5 \cdot J_z$, whereas in the right one we assume $\sigma_{J_z}= 1 \cdot J_z$.    }
\label{ES20_zmax_mg_error}
\end{centering}
\end{figure}

\begin{figure*}
\begin{centering}
\includegraphics[scale=0.42]{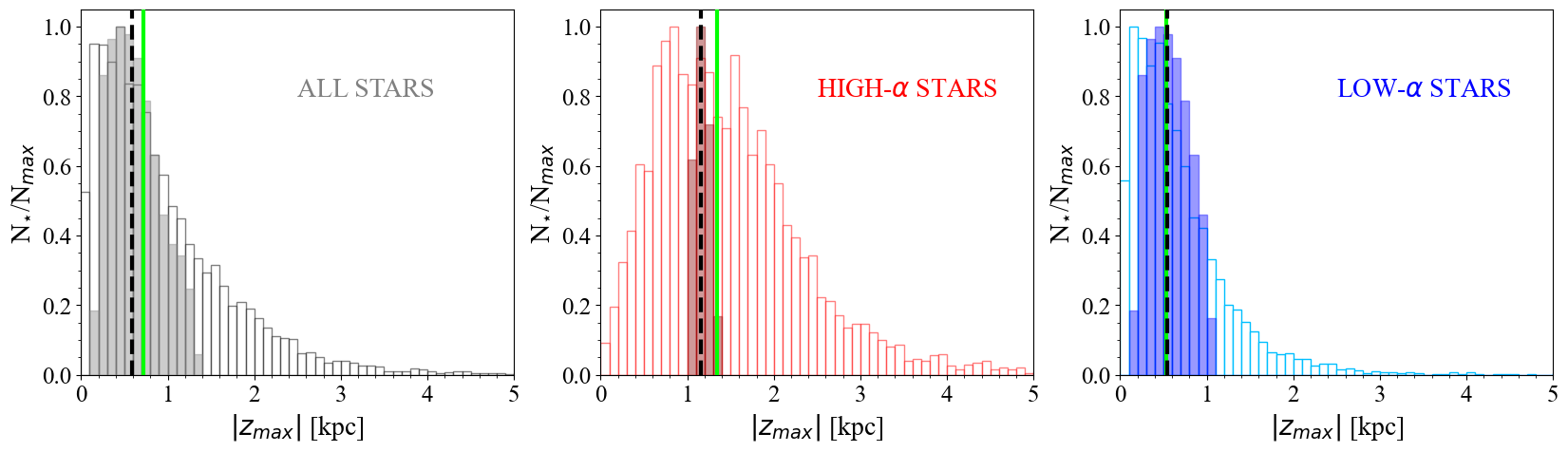}
\includegraphics[scale=0.42]{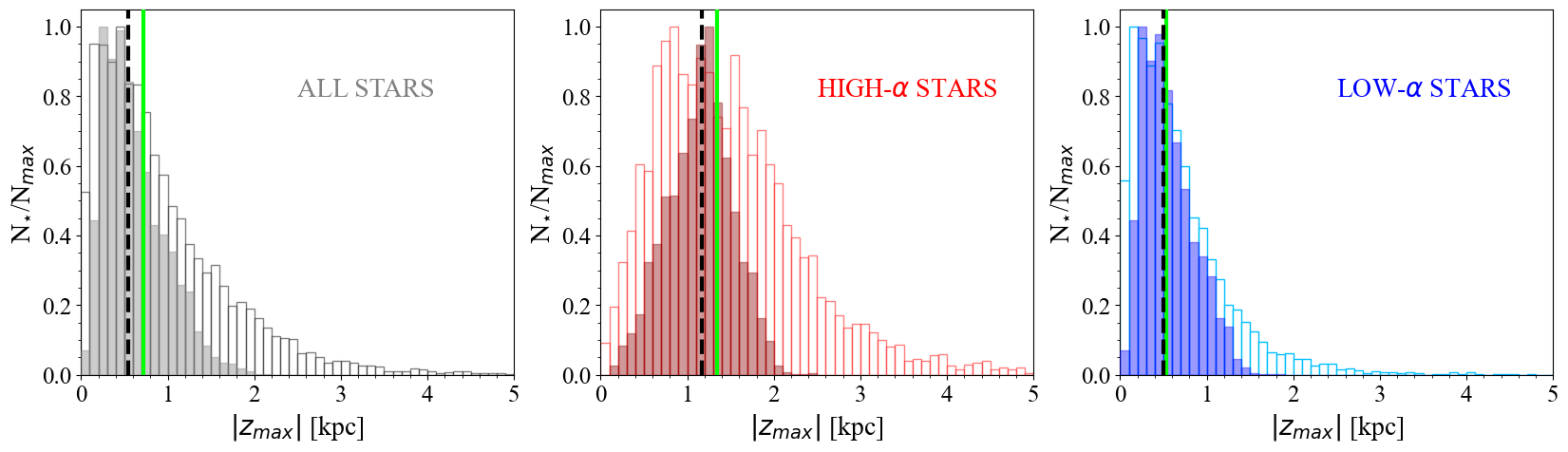}
\includegraphics[scale=0.42]{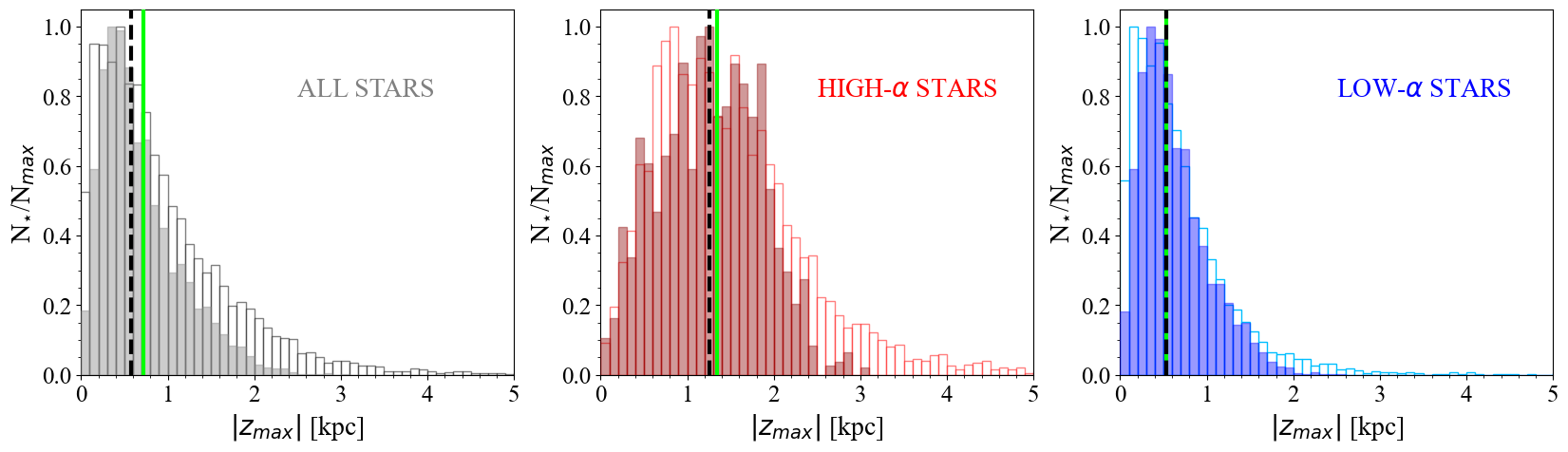}
\caption{ As in Fig. \ref{ES21_distrib} but for the ES20 model. }
\label{ES20_distrib}
\end{centering}
\end{figure*}
\citet{gandhi2019}, studying the stellar dynamics for   LAMOST stars with distances smaller than 2 kpc from the Sun, found a large dispersion in the computed  vertical action  $J_z$  versus age relation, i.e. for the high-$\alpha$ sequence they find   $\sigma_{J_z}/J_z=0.9$ and   $\sigma_{J_z}/J_z=1.13$ for low-$\alpha$ stars. Similar results are found by \citet{ting}.

We take into account this dispersion in our model by adding, at each Galactic time $t$, a random error which has a Gaussian distribution to the  vertical action $\widehat{J_z}$ of \citet{ting}  associated to the SSP 
formed at Galactic time $t$. 

The   ``new'' vertical action  $\widehat{J}_{z, \mbox{ new}}(t)$ is defined as:

\begin{equation} 
 \widehat{J}_{z, \mbox{ new}}(t)   = \widehat{J_z} (t)  +  \delta_G(\widehat{J_z} (t));  \, \mbox{  } \delta_G(t)  \sim  \mathcal{N}(0, \sigma_{J_z}),
\label{sigma_J}
\end{equation}
 where $\delta_G$ is a perturbation which follows a Normal distribution $\mathcal{N}(0, \sigma_{J_z})$ with standard deviation $\sigma_{J_z}$. 
In conclusion, we  perturbed  the $J_z$-age relation of \citet{ting}, and  $\widehat{J}_{z, \mbox{ new}}$ is taken as  the new constraint when we  compute  $J_z$ with \textsc{galpy} varying the initial vertical velocity  as indicated in Eq. (\ref{vo_ref}).

In the light of \citet{gandhi2019} and \citet{ting} results, in Fig. \ref{jz_age_disper} we show the $J_z$ versus age relations and the initial vertical velocities $v_{0, z}(\tau)$ for the SSPs  in the solar neighbourhood with $\sigma_{J_z}/J_z=0.5$  and  $\sigma_{J_z}/J_z=1$. 
In Fig. \ref{ES21_distrib}, we also show the \zmax  distributions with $\sigma_{J_z}/J_z=0.5$ (second row) and  $\sigma_{J_z}/J_z=1$ (third row).
We note  that the predicted medians of the \zmax distribution  with $\sigma_{J_z}/J_z=1$   for both the high- and low- $\alpha$ sequences   are in good agreement with the data. Moreover, the observed spread of the distribution is satisfactory reproduced. On the other hand,  the $\sigma_{J_z}/J_z=0.5$ case (as reported in the second row)  does not allow to mimic the observed spread in the high-$\alpha$ sequence.  
Actually, also with $\sigma_{J_z}/J_z=1$  the high \zmax tail of the distribution (stars with \zmax> 3 kpc) cannot be  predicted by the model. We think that other sources of errors could be the origin of this discrepancy. However, here we want to focus solely on the effects of uncertainties on the $J_z$ determination.

In Fig. \ref{ageES21}, we also show the predicted age versus \zmax distributions assuming $\sigma_{J_z}/J_z=0.5$ (left panel) and  $\sigma_{J_z}/J_z=1$ (right panel) compared to astroNN ages   (computed by \citealt{mackereth2019} using a Bayesian neural network model trained on asteroseismic ages). We can appreciate that the trend of  observed spread in the \zmax versus age distribution   is better reproduced with $\sigma_{J_z}/J_z=1$.
In the lower panels of  Fig. \ref{ageES21} we can appreciate that the distributions of the predicted SSPs in the  \zmax versus [Mg/Fe] space including the observed dispersion in the vertical action  $J_z$ estimates, highlight the presence of the disc dichotomy signature, in agreement with data.

As already noticed in ES21, the predicted  distribution of surface stellar mass density $\Delta M_{\star}$ (computed in constant age intervals fixed at the value  of 0.05 Gyr) in the [Mg/Fe] versus [Fe/H] relation is in contrast with APOGEE DR16 stellar distribution. Also in  Fig. \ref{ES21all}, the model forms too many stars as soon the first infall begins, and in the [Mg/Fe] vs \zmax they populate a region where APOGEE DR16 are rare,  i.e. [Mg/Fe] with values larger than 0.3 dex. This is due to the fact that the best fit-model parameter value for the infall time-scale $\tau_1$ of the high-$\alpha$ sequence is $\sim 0.1$ Gyr, hence the bulk of high-$\alpha$ stars are created in  region in the \zmax versus  [Mg/Fe]
plane, where population density is small in the APOGEE DR16. In the Section \ref{s:ES20} we will show  that a  longer time-scale of gas accretion $t_1$ in the high-$\alpha$ sequence which  characterised the ES20 model (see Table \ref{tab_mcmc}), is able to alleviate this tension.

\subsection{ES20 model}
\label{s:ES20}
In this Section, we compare the combined  APOGEE data + astroNN catalogue with model predictions based on the chemical evolution model \emph{M2} of ES20.
In Fig. \ref{ES20all} we show model results with extended  \citet{ting} relation to high-$\alpha$ sequence stars in computing the maximum height \zmax.
As for the ES21
model (see Section \ref{s:ES21}), a better agreement with  the data  is obtained when  APOGEE DR16 stars with observed vertical heights $|z|<2$ kpc are considered.

As we pointed out in Section \ref{s:ES21},  there is a discrepancy between the most  densely populated regions in the   maximum vertical heights versus [Mg/Fe]  by APOGEE DR16 stars and the  peaks of  the predicted distribution of the stellar mass density found by ES21 model.
On the other hand, in Fig. \ref{ES20all} we  note  that the distribution   of  the formed stars predicted by ES20 model throughout the curve in the   \zmax versus [Mg/Fe] space shows no tensions with the data.   In fact, the two peaks of the predicted stellar mass density  are  in correspondence to the highest data density regions as traced by the contour density curves.
We recall that the best-fit parameters in model \emph{M2} of ES20  have been constrained by the APOKASC sample,  taking into account  also stellar ages computed with asteroseismology in the MCMC calculations. 
As discussed in Section \ref{s:chemmod}, longer  timescales $t_1$ and $t_2$ for  the  gas  accretion are predicted by this model constrained by APOKASC sample (see Table \ref{tab_mcmc}). Regarding the high-$\alpha$ sequence, the main consequence of the slower gas accretion is to delay the peak of the star formation towards smaller [Mg/Fe] values.

In Fig. \ref{dm} we report the  predicted surface stellar mass density $\Delta M_{\star}$ formed in age intervals of $0.05$ Gyr as function of the [Mg/Fe] abundance ratio predicted by  ES20 and   ES21 models, respectively.
Here, it is even more evident that in the ES20 model the peak of the stellar mass formed during the high-$\alpha$ sequence phase is shifted towards  smaller [Mg/Fe] values.

 In   Fig. \ref{ES20_zmax_mg_error} we notice  that the distributions of the predicted SSPs in the  \zmax versus [Mg/Fe] space including  in the ES20 model the observed dispersion in the vertical action  $J_z$ estimates, show  the presence of the disc dichotomy feature, in good agreement with data.

 Finally, in Fig. \ref{ES20_distrib} we show that also for the ES20 model the median values  of the observed \zmax distributions of APOGEE-DR16+astroNN stars   are well reproduced for both the high- and low- $\alpha$ sequences. However, as already discussed in Section \ref{s:ES21}, in order to mimic the observed spread we have to include in the model the errors of $\sigma_{J_z}= 1 \cdot J_z$ in the considered vertical action  $J_z$   (see the last row of Fig. \ref{ES20_distrib}).

\begin{figure}
\begin{centering}
\includegraphics[scale=0.4]{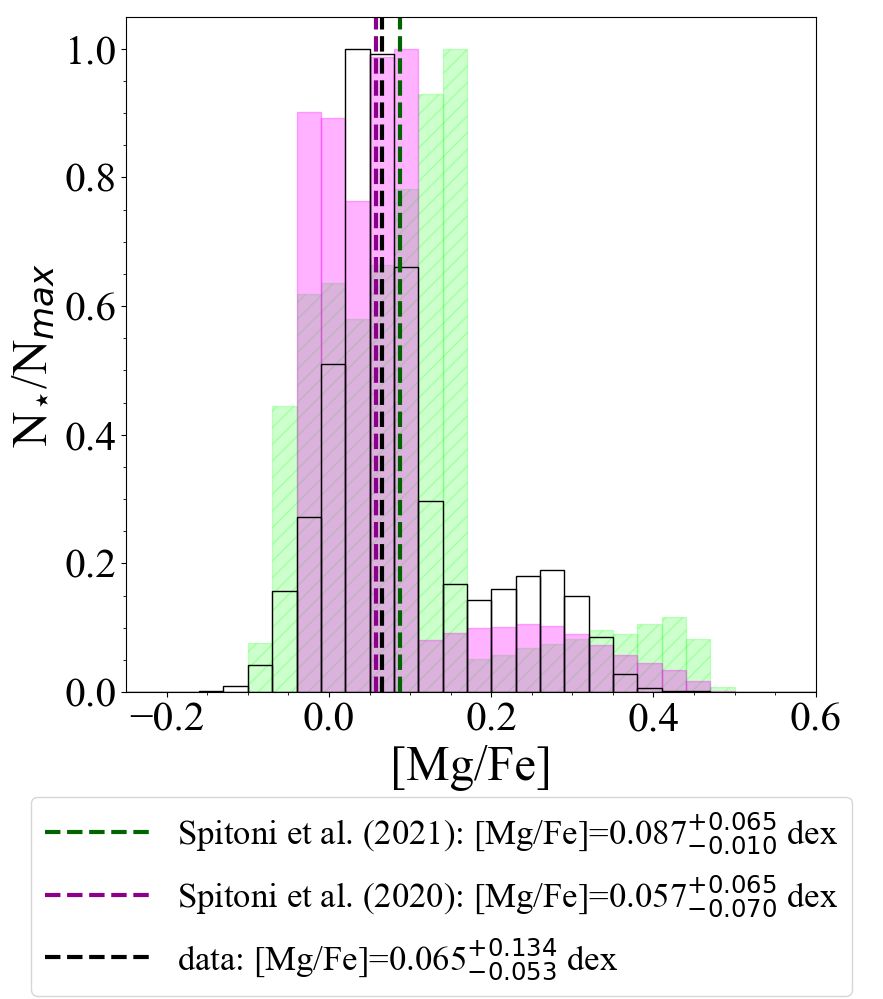}
\caption{[Mg/Fe] distributions  predicted by ES21 model computed at 8 kpc (green histogram) and by model \emph{M2} of ES20 (magenta histogram)  compared with the APOGEE DR16 data (black empty histogram) for stars with Galactocentric distances between 6 and 10 kpc. Black, green and magenta vertical dashed lines indicate the median values of the data and models. }
\label{distrib}
\end{centering}
\end{figure}

\begin{figure*}[h]
\begin{centering}
\includegraphics[scale=0.35]{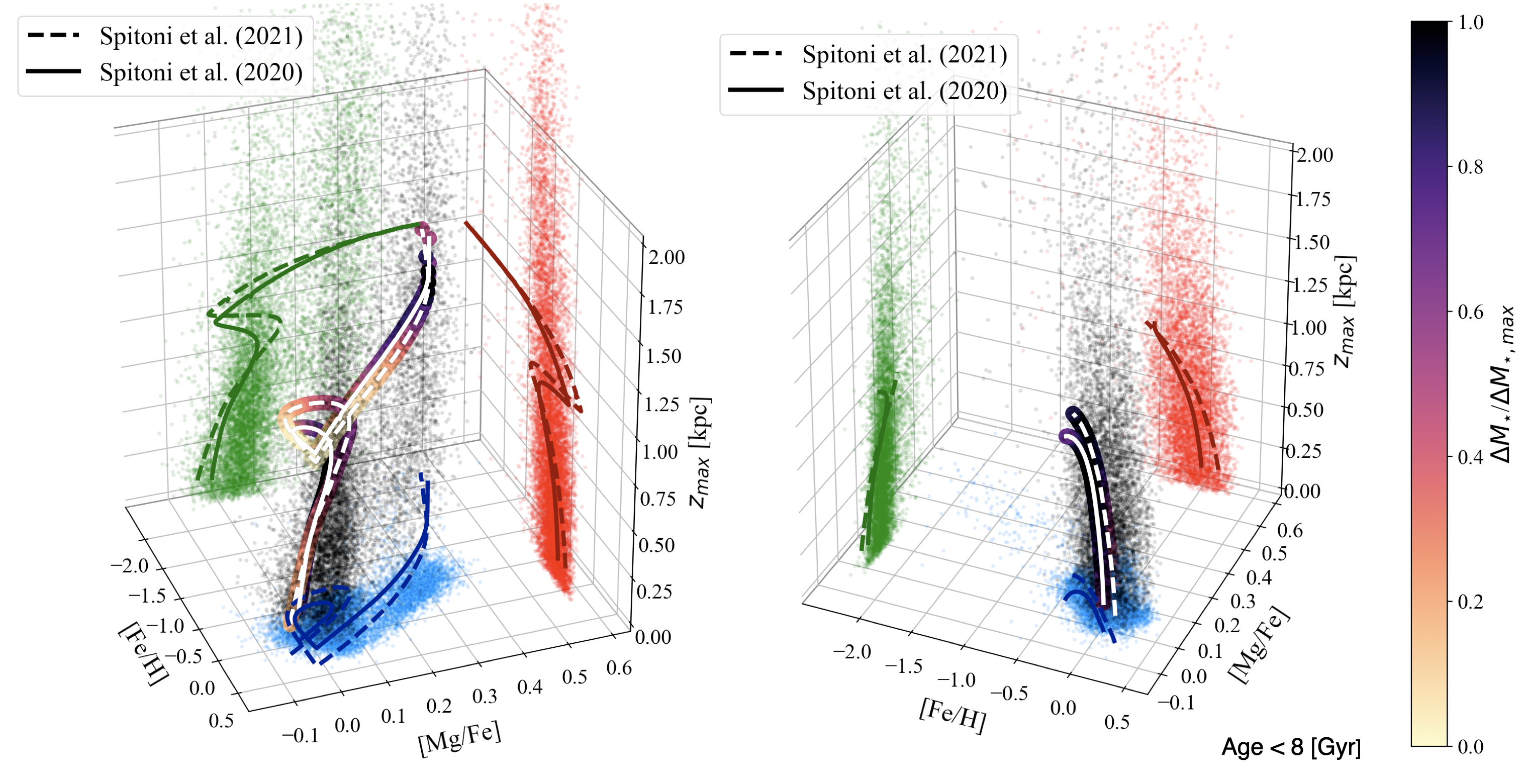}
\caption{Observed stars (grey points) in the 3D space formed by  APOGEE DR16 abundance ratios (i.e. [Fe/H] and   [Mg/Fe]) and maximum vertical excursion from the midplane   \zmax (from astroNN catalogue)  in the Galactic region between 6 and 10 kpc with $|z|<$ 2 kpc  compared with model predictions by ES20 and  ES21 (white solid and dashed lines, respectively). The associated projections are drawn with red, blue, green points (data) and lines (models), respectively. In the left panel all the stellar ages have been considered, while in the right one only observed and predicted stars  younger than 8 Gyr have been drawn. In both panels, the colour coded points indicate the predicted surface stellar mass density $\Delta M_{\star}$ formed in constant age intervals fixed at the value of 0.05 Gyr normalised to the maximum value $\Delta M_{\star, max}$.   }
\label{3d}
\end{centering}
\end{figure*}

\subsection{[Mg/Fe] distribution functions and 3D space  ([Mg/Fe], [Fe/H], \zmax) analysis}

In Section \ref{s:ES21}, we pointed out  that the  predicted distribution in the \zmax  versus  [Mg/Fe]  relation by  ES21  of  the newly formed  stellar mass in the same age bins  (i.e., $\Delta \tau=0.05$ Gyr) in the high-$\alpha$ sequence      is in contrast  with the APOGEE data. On the other hand, this tension is resolved when adopting the ES20 model.
In Fig. \ref{distrib}, we compare the [Mg/Fe] distribution function computed by ES21 and  ES20 models with the APOGEE DR16 data  in the annular Galactic region enclosed between 6 and 10 kpc (to be consistent with the distribution reported in   Fig. 16 of \citealt{spitoni2021} we consider  stars with vertical heights $|z|<1$ kpc).   ES21  best-fit model (green histogram) accounts for the observed bimodality, however the predicted peaks are significantly shifted towards higher [Mg/Fe] values. As underlined in Section \ref{s:ES21}, this is due to the short timescale of accretion $\tau_1$ and $\tau_2$ as obtained by the MCMC calculation when only chemical APOGEE DR16 abundances are fitted, which push the star formation activity   to be concentrated close to the peaks of maximum    gas infall rates.  
On the contrary, model \emph{M2} by ES20 characterised by  longer time-scales of accretion (see Table \ref{tab_mcmc}) predicts the above-mentioned bimodality and  is capable to  reproduce the [Mg/Fe] values  of   high-$\alpha$ and low-$\alpha$ distribution peaks as shown in APOGEE DR16 data. We conclude that the inclusion of precise stellar ages inferred from asteroseismology is  fundamental in order to properly  constrain  chemical evolution models of the Milky Way disc components. 

Finally, in Fig. \ref{3d} we compare models predictions in the 3D space formed by  the abundance ratios  [Mg/Fe], [Fe/H]  and  the orbital parameter \zmax  with APOGEE DR16 data (observed Galactic heights $|z|<2$ kpc). In the left panel of Fig. \ref{3d} we analyse   the whole stellar ages range.
Colour-coded points  indicating the stellar mass $\Delta M_{\star}$ formed in age intervals of 0.05 Gyr,    highlight  that in  ES21 model the star formation in the high-$\alpha$ sequence is concentrated towards the infall rate peak (see also Fig. \ref{dm}), whereas in  ES20 it is characterised by a more extended star formation history and as mentioned above the predicted [Mg/Fe] distribution function is in agreement with APOGEE DR16. 
In the projections, it can be noted that the evident dichotomy present in the data in the plane \zmax versus  [Mg/Fe] is not so clear in   \zmax versus [Fe/H].
On the right panel of Fig.  \ref{3d}, where we consider only stars younger than 8 Gyr  it is possible to  appreciate a good agreement between model predictions and data.

\section{Conclusions}
\label{s:conclusions} 
Recent chemical evolution models designed to reproduce APOGEE DR16 data \citep{spitoni2021} and APOKASC sample \citep{spitoni2020}  suggested the presence of   a significant delay time  between the two gas infall episodes  for the thick-disc and thin-disc formation.
In this work, we presented results for 
 the vertical distribution of the [Mg/Fe] abundance ratio  in the solar neighborhood and showed how this is consistent with recent  observations combining the APOGEE DR16 data (chemical abundances) and the astroNN catalogue (stellar ages, orbital parameters).   We computed the vertical maximum heights \zmax   using the \citet{ting} relation in computing the orbits around the Galaxy  of SSPs born at different evolutionary times.
Our main conclusions can be summarised as follows:
\begin{enumerate}

\item  Regarding the vertical [Mg/Fe] abundance distributions,  we have a  better agreement between   models \citet{spitoni2020,spitoni2021} and combined APOGEE DR16 data and astroNN catalogue 
(for stellar ages younger than 8 Gyr)   for stars close to the Galactic mid-plane $|z|<2$~kpc.

 \item The distribution of the  initial vertical velocities $v_{0,z}(\tau)$  as the function of the Galactic age $\tau$ for the computed SSPs can be interpreted as the upper limit of the observed   vertical velocity dispersion $\sigma_z$  versus age relation in the solar vicinity.

\item Extending \cite{ting} relation to the whole stellar age range the predicted curves for the  vertical  [Mg/Fe] distribution  by the models nicely overlap with the regions characterised by the highest densities of stars in both low-$\alpha$ and high-$\alpha$ APOGEE sequences with observed heights $|z|<2$ kpc.
 However, the median values of  APOGEE data in the high-$\alpha$ sequence  generally show a steeper growth of  \zmax compared to model predictions. This is due to the fact that in the past the Milky Way could have been  affected by important merger episodes, and such as external perturbations could  have  heated up the thick disc. Hence, an extra-vertical action $J_z$ component  is missing if  we   consider stellar scattering  as the only  heating process  as assumed by \citet{ting}.

\item Assuming  for APOGEE-DR16  stars ($|z|<2$~kpc) a  disc dissection  based on chemistry,  the  observed  $|z_{max}|$ distributions for  high-$\alpha$ and low-$\alpha$ sequences are in good agreement with our model predictions if we consider in the calculation an error in  the vertical action $\widehat{J_z}$ of \citet{ting} with standard deviation $\sigma_{J_z}=1 \cdot J_z$ (such  an error has been inspired  by the study of \citealt{gandhi2019} based on GALAH stars and \citealt{ting}).

\item   When we also include  the information  about  the predicted surface stellar mass density throughout the chemical evolution in  \citet{spitoni2021} model results,  there is a tension between the location of  most  densely populated regions in  APOGEE DR16 stars and the   model peaks in the \zmax versus [Mg/Fe]  relation if we consider the full  stellar age range.    In fact, \citet{spitoni2021} model forms too many stars as soon the first infall begins, and in the  \zmax versus [Mg/Fe] many stars are predicted in a region where APOGEE DR16 stars are rare.  This is the consequence of  the  best fit-model parameter value for the infall time-scale $\tau_1$ of the high-$\alpha$ sequence is quite short ( i.e,  $\sim 0.1$ Gyr).
 \item On the contrary, model \emph{M2} by \citet{spitoni2020} characterised by  longer time-scales of accretion (see Table \ref{tab_mcmc}) reproduces   the above-mentioned bimodality  in the  \zmax versus [Mg/Fe] relation. 
 The distribution of the formed stars predicted by \citet{spitoni2020} model in the  \zmax versus [Mg/Fe] relation shows no tensions with the data.  In  fact,  the  two  peaks  of  the  predicted  stellar  mass  density  are    in  correspondence  to  the  highest   density regions in the data.
 We conclude that the inclusion of precise stellar ages inferred from asteroseismology is  fundamental to properly  constrain  chemical evolution models of the Milky Way disc components.

 \item  The distributions of the predicted SSPs in the  \zmax versus [Mg/Fe] space including the observed dispersion in the vertical action  $J_z$ estimates, show  the presence of the disc dichotomy signature, in good agreement with data.   
\end{enumerate}

\section*{Acknowledgement}
 The authors thank the referee F. Vincenzo for various suggestions that improved the paper. E. Spitoni  thanks A. Recio-Blanco and P. A. Palicio for useful discussions. 
E. Spitoni received funding from the European Union’s Horizon 2020 research and innovation program under SPACE-H2020 grant agreement number 
101004214 (EXPLORE project).
E. Spitoni  acknowledges support from  the ERC Consolidator Grant (Hungary) programme (Project RADIOSTAR, G.A. n. 724560).
Funding for the Stellar Astrophysics Centre is provided by The Danish National Research Foundation (Grant agreement no.: DNRF106). E. Spitoni and V. Aguirre B\o rsen-Koch acknowledge support from the Independent Research Fund Denmark (Research grant 7027-00096B).
 This work was partially supported by the program Unidad de Excelencia María de Maeztu CEX2020-001058-M. K. Verma is supported by the Juan de la Cierva fellowship (IJC2019-041344-I).
A. Stokholm acknowledges support from the European Research Council Consolidator Grant funding scheme (project ASTEROCHRONOMETRY, G.A. n. 772293, \url{http://www.asterochronometry.eu}).

 In this work, we have made use of SDSS-IV APOGEE-2 DR16 data. Funding for the Sloan Digital Sky Survey IV has been provided by the Alfred P. Sloan Foundation, the U.S. Department of Energy Office of Science, and the Participating Institutions. SDSS-IV acknowledges
support and resources from the Center for High-Performance Computing at
the University of Utah. The SDSS web site is  \href{www.sdss.org}{www.sdss.org}.
SDSS is managed by the Astrophysical Research Consortium for the Participating Institutions of the SDSS Collaboration which are listed at \href{https://www.sdss.org/collaboration/affiliations/}{www.sdss.org/collaboration/affiliations/}. With this paper we also made use of the Python package for Galactic dynamics \textsc{GALPY} (\href{http://github.com/jobovy/galpy}{http://github.com/jobovy/galpy}).
\bibliographystyle{aa} 
\bibliography{disk}

\end{document}